\documentclass[12pt,headings=big,numbers=noenddot,DIV=14,a4paper]{article}%

\pdfoutput=1

\usepackage[margin=1.25in]{geometry}

\usepackage{amsmath,amssymb,amsfonts}
\usepackage[normal,font=small,labelfont=bf,labelsep=period]{caption}
\usepackage[pdftex]{color,graphicx}
\usepackage[english]{babel}
\usepackage[compress]{cite}

\usepackage[dvipsnames]{xcolor}
\definecolor{darkgreen}{rgb}{0,0.65,0}
\usepackage{slashed}

\usepackage[linktoc=page,bookmarks=false,colorlinks=true,
linkcolor=WildStrawberry,citecolor=BlueGreen,urlcolor=RedViolet]{hyperref}

\numberwithin{equation}{section}
\DeclareFontFamily{OT1}{pzc}{}
\DeclareFontShape{OT1}{pzc}{m}{it}{<-> s * [1.10] pzcmi7t}{}
\DeclareMathAlphabet{\mathpzc}{OT1}{pzc}{m}{it}

\newcommand{\beq}{\begin{equation}}
\newcommand{\eeq}{\end{equation}}

\newcommand{\mDM}{m_\text{DM}}

\def\LHC{{\sc LHC}}

\def\FERMI{{\sc Fermi}} 
\def\HESS{{\sc H.E.S.S.}}

\def\HESSII{{\sc H.E.S.S.-II}}
\def\MAGIC{{\sc MAGIC}}
\def\VERITAS{{\sc VERITAS}}
\def\PAMELA{{\sc PAMELA}}
\def\CTA{{\sc CTA}}
\def\AMSII{{\sc AMS-02}}
\def\LZ{{\sc LZ}}
\def\XenonT{{\sc XENON-1T}}
\def\Darwin{{\sc DARWIN}}

\def\Dra{{\tt  Draco}}
\def\Tri{{\tt  Triangulum-II}}
\def\Scu{{\tt  Sculptor}}
\def\Seg{{\tt  Segue~I}}

\definecolor{rosso}{cmyk}{0,1,1,0.4}

\definecolor{blue}{rgb}{0.1,0.1,0.9}

\begin{document}

\begin{flushright}
\footnotesize
\end{flushright}
\color{black}

\begin{center}

{\huge \bf 
Dark Matter in $\gamma$ lines: \\[.1cm]
Galactic Center vs dwarf galaxies
}

\medskip
\bigskip\color{black}\vspace{0.5cm}

{
{\large\bf Valentin Lefranc}$^a$,
{\large\bf Emmanuel Moulin}$^a$,
{\large\bf Paolo Panci}$^b$, \\[.1cm]
{\large\bf Filippo Sala}$^c$,
{\large\bf Joseph Silk}$^{b,d,e}$
}
\\[7mm]

{\it $^a$ DRF/Irfu, Service de Physique des Particules, CEA Saclay, F-91191 Gif-Sur-Yvette Cedex, France}\\[3mm]
{\it $^b$ Institut d'Astrophysique de Paris, UMR 7095 CNRS, Universit\'e Pierre et Marie Curie,
98 bis Boulevard Arago, Paris 75014, France}\\[3mm]
{\it $^c$ LPTHE, UMR 7589 CNRS, 4 Place Jussieu, F-75252, Paris, France}\\[3mm]
{\it $^d$ The Johns Hopkins University, Department of Physics and Astronomy,
3400 N. Charles Street, Baltimore, Maryland 21218, USA}\\[3mm]
{\it $^e$ Beecroft Institute of Particle Astrophysics and Cosmology, Department of Physics, University of Oxford, 1 Keble Road, Oxford OX1 3RH, UK}\\[3mm]
\end{center}

\bigskip

\centerline{\bf Abstract}
\begin{quote}
\color{black}

We provide \CTA\ sensitivities to Dark Matter (DM) annihilation in $\gamma$-ray lines, from the observation of the Galactic Center (GC) as well as, for the first time, of dwarf Spheroidal galaxies (dSphs).
We compare the GC reach with that of dSphs as a function of a putative core radius of the DM distribution, which is itself poorly known. We find that the currently best dSph candidates constitute a more promising target than the GC, for core radii of one to a few kpc.
We use the most recent instrument response functions and background estimations by \CTA, on top of which we add the diffuse photon component.
Our analysis is of particular interest for TeV-scale electroweak multiplets as DM candidates, such as  the supersymmetric Wino and the Minimal Dark Matter fiveplet, whose predictions we compare with our projected sensitivities.
\end{quote}

\newpage
\tableofcontents

\section{Motivation}

By revealing the existence of the Higgs boson, the Large Hadron Collider (\LHC) has now completed the Standard Model of particle physics (SM). However, the \LHC\ has not found clear signs of new physics (NP), casting doubts on the so-called \textit{naturalness} of the electroweak (EW) scale, which is arguably the strongest theoretical argument to expect new physics to exist in that range.
Still, the simple assumption of Dark Matter (DM) being a weakly interacting massive particle (WIMP) motivates NP close to the electroweak scale, independently of naturalness arguments, thanks to the theoretically attractive requirement of obtaining the correct relic DM abundance via thermal freeze-out.
The quest for such DM candidates is then a crucial motivation of the present experimental exploration at the energy and astroparticle frontiers.

The hitherto empty-handed experimental searches (at colliders and in direct and indirect detection) for such WIMPs are now pushing, quite generically, the mass of this kind of DM to the TeV scale.
Such large masses pose a challenge to collider experiments, because of the corresponding center-of-mass energy needed to produce DM, as well as to direct detection experiments, because of the lower number density of DM particles that correspond to larger DM masses. The weak DM interactions further undermine the sensitivity of both experiments to its signals.

Indirect detection experiments are thus essential in order  to probe these regimes. In their regard, lower DM particle densities imply dimmer cosmic rays fluxes originating from DM annihilations. This motivates the quest for spectral features, like lines, in the cosmic rays (CR) spectra, since they are readily distinguishable from astrophysical backgrounds even if their fluxes are moderate. 
Moreover, higher DM masses imply that the corresponding CR fluxes are expected at higher energies, where the sensitivity of some telescopes -typically satellites, such as \PAMELA, \AMSII\ and \FERMI- is limited by their size.  This favours telescopes with a large effective area, most notably ground-based Cherenkov arrays, and puts them in an excellent situation to explore the WIMP Dark Matter paradigm.
Cherenkov telescopes measure high-energy (up to $\sim$ 100 TeV) cosmic $\gamma$-rays from specific sky regions. Currently \HESS~\cite{hess}, \MAGIC~\cite{magic} and \VERITAS~\cite{veritas} are operating, and the next-generation Cherenkov Telescope Array (\CTA)~\cite{Consortium:2010bc} is expected to start taking data towards the end of this decade.

From the point of view of particle physics, the DM annihilation in $\gamma$-rays is generically loop-suppressed with respect to annihilations into other cosmic rays. However, the so-called Sommerfeld effect partially compensates this hierarchy, for WIMPs with masses sufficiently larger than the mediators of their interactions with the SM. 
This renders $\gamma$-ray lines one of the most promising probes, a prominent example being that of multi-TeV scale WIMPs whose interactions are mediated by the SM EW gauge bosons.

From a more astrophysical point of view, the fact that the DM annihilation signals scale as $\rho_{\rm DM}^2$ makes it interesting to explore regions where the DM density profile $\rho_{\rm DM}$ is enhanced, like the Galactic Center (GC) and dwarf Spheroidal galaxies (dSphs).
The first has already been widely observed in the quest for lines~\cite{Abramowski:2013ax,Ackermann:2015lka,Kieffer:2015nsa}, and has allowed  placing stringent constraints on WIMP models, despite the large baryonic content and the poor knowledge of $\rho_{\rm DM}$ in the inner galactic regions.
DSphs are much cleaner from baryons, being the most DM dominated objects known, but only Segue-I has so far been searched for $\gamma$ lines~\cite{Aleksic:2013xea}. The high rate at which new dSphs are being discovered, 
with excellent prospects for indirect detection, further encourages their exploration in terms of high energy $\gamma$-ray lines. 

In this paper, motivated by the above considerations, we study the sensitivity of \CTA\ to $\gamma$ lines from both the GC and dSphs.
We provide the expected reach in DM annihilation cross-sections into $\gamma \gamma$, and we also compare our predictions with some prototypical WIMP models (fermion multiplets of the SM gauge group $SU(2)_{\rm L}$) as benchmark targets. 
We discuss the interplay of the GC and dSph reaches, in light of the radius of the possible core of the DM distribution in the Milky Way (MW) halo.

The exposition is organised as follows: in sec. \ref{sec:multiplets} we review EW multiplets and their phenomenology, including the Sommerfeld enhancement; in sec. \ref{sec:gammas} we discuss the signals that we look for, in terms both of their spectral features and of the observation targets (the GC and dSphs); in sec.~\ref{sec:CTA} we describe the derivation of the \CTA\ projected sensitivities to $\gamma$-ray lines; in sec.~\ref{sec:results} we present our results; in sec.~\ref{sec:conclusion} we conclude.

\section{Electroweak multiplets as DM candidates}
\label{sec:multiplets}

%
Multiplets of the SM gauge group $SU(2)_{\rm L}$ are, by definition, the prototype of WIMP DM, and therefore constitute a motivated and economical benchmark for phenomenological studies. Besides being simple benchmarks, fermion $SU(2)_{\rm L}$ multiplets with zero hypercharge arise as DM candidates in motivated NP constructions.
We focus here on the cases of a Majorana triplet (also known as ``Wino'') and fiveplet as particularly motivated examples:
\begin{itemize}
\item[$\diamond$] A \emph{Wino} is predicted to be the lightest supersymmetric particle (LSP), and therefore a candidate of DM, in models of anomaly mediation~\cite{Giudice:1998xp}. It is further motivated in models of Supersymmetry with heavy scalars~\cite{Wells:2003tf}, from the first split-SUSY proposals~\cite{ArkaniHamed:2004fb,Giudice:2004tc} up to more recent constructions inspired by the anthropic principle (see~\cite{D'Eramo:2014rna} and references therein). 
Irrespective of Supersymmetry, an EW fermion triplet marries minimality with other virtues, see \textit{e.g.} the discussion in~\cite{Cirelli:2014dsa}.
\item[$\diamond$] The \emph{MDM 5plet} emerges as the only~\cite{DiLuzio:2015oha,DelNobile:2015bqo} viable DM candidate in the Minimal Dark Matter (MDM) framework~\cite{Cirelli:2005uq,Cirelli:2009uv}, where the SM is enlarged by only one particle, demanded to be (accidentally) stable thanks to gauge invariance. 
\end{itemize}

\subsection{Phenomenological status}

The phenomenology of such candidates is controlled by only one free parameter, $M_{\rm DM}$, which is fixed to 2.9~TeV (triplet)~\cite{Beneke:2016ync} and 9.4~TeV (fiveplet)~\cite{Cirelli:2015bda} if the thermal freeze-out requirement is imposed. The neutral component of the multiplet is the DM, and its coannihilations with the other components in the early universe effectively increase its annihilation cross-section, and consequently the mass for which the measured relic abundance is achieved, with the respect to the naive WIMP picture. These masses are further pushed into the multi-TeV range by the Sommerfeld enhancement of the annihilation cross-section~\cite{Hisano:2006nn}. 
In the rest of this discussion, we will leave $M_{\rm DM}$ as a free parameter: on the one hand, this provides a good quantification of the experimental reach, on the other hand, DM production could well be non-thermal, so that the DM mass could either be lighter (see \textit{e.g.}~\cite{Moroi:1999zb}) or heavier (see \textit{e.g.}~\cite{Giudice:2000ex}) than the aforementioned thermal values.

Coming now to experimental probes of these candidates, the LHC currently excludes masses lighter than $\sim 270$ GeV~\cite{Aad:2013yna,CMS:2014gxa}, and TeV masses are not expected to be reached before construction of a 100 TeV collider, which has the potential to probe a thermal Wino~\cite{Low:2014cba,Cirelli:2014dsa}, and a fiveplet with a mass of $\sim 5$ TeV~\cite{Golling:2016gvc}.
Concerning direct detection, the spin-independent cross-section with nuclei of these candidates is accidentally suppressed~\cite{Hisano:2015rsa}, so that TeV masses are out of reach at current and near future experiments, up to \XenonT~\cite{Cushman:2013zza}. The TeV territory will be explored by \LZ, and the thermal masses are expected to be probed only with multi-ton future detectors, such as  \Darwin~\cite{Darwin:2016}.
Concerning indirect detection, $\gamma$-rays from the GC (as first emphasised in \cite{Cohen:2013ama,Fan:2013faa} for lines) and dSph are, at present, the most promising probes~\cite{Hryczuk:2014hpa,Cirelli:2015bda}. Indeed, \HESS\ observations of $\gamma$-ray lines from the GC~\cite{Abramowski:2013ax} already exclude both thermal benchmarks, if a profile like Einasto or NFW is assumed. For more cored profiles instead, the masses excluded are lower than the thermal ones, and whether the GC center would be a more promising target, with respect to dwarves, is at present not clear - and partly motivates this work. The crucial ingredient of indirect detection of EW multiplets is the Sommerfeld enhancement, which we review next.

\subsection{Sommerfeld-enhanced cross sections}

\begin{figure}[!t]
\centering
\includegraphics[width=0.495\textwidth]{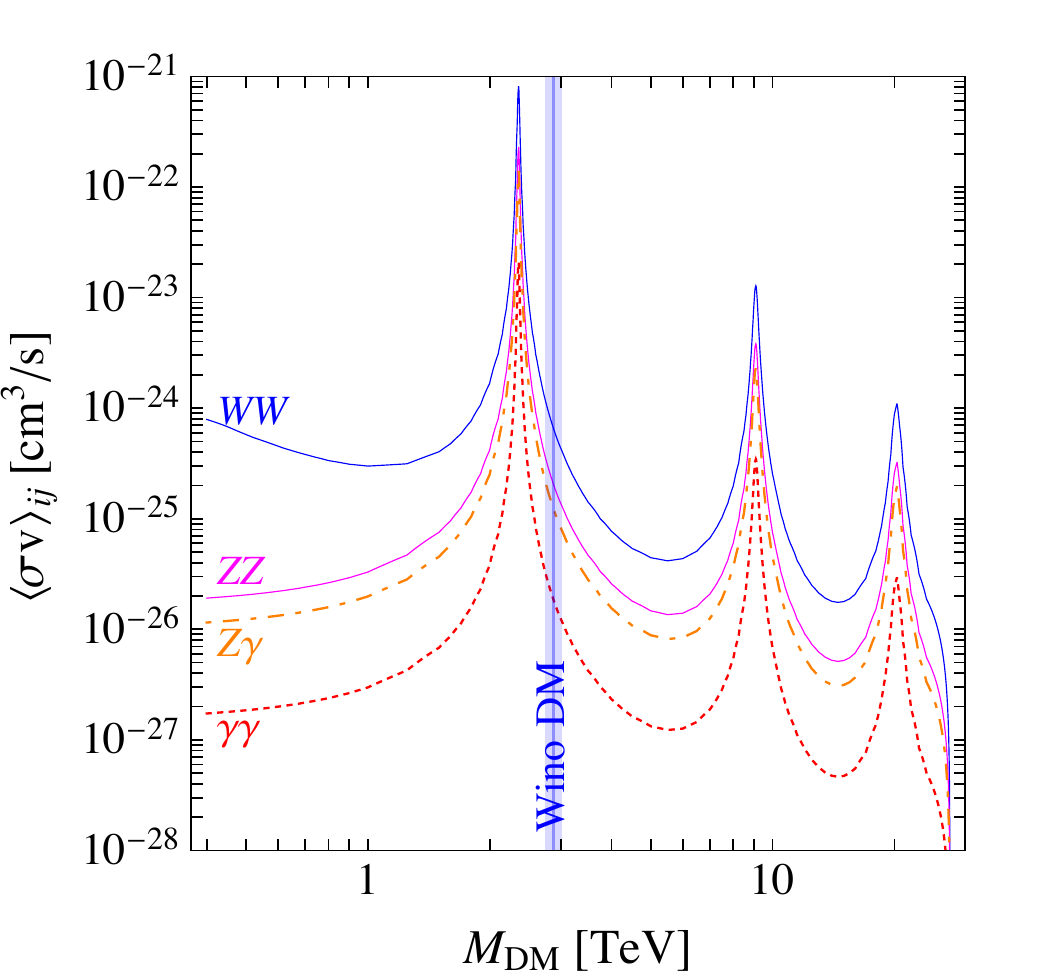} 
\includegraphics[width=0.495\textwidth]{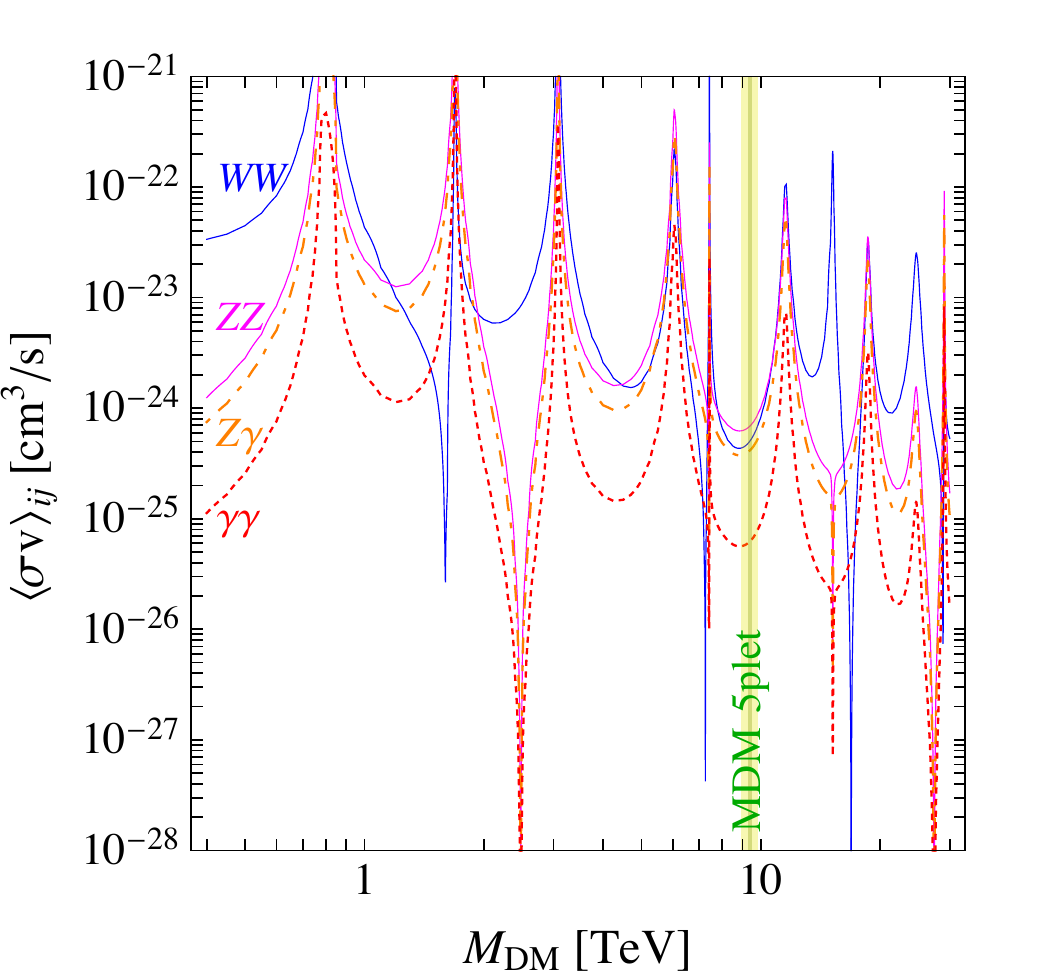} 
\caption{
\em \small \label{fig:NR_Sommerfeld_XS} {\bfseries Annihilation cross-sections into SM gauge bosons.} Left: Wino DM. Right: MDM 5plet. $W^+W^-$, $ZZ$, $Z\gamma$ and $\gamma\gamma$ are shown respectively with blue, magenta dashed, orange dot-dashed, and red dotted lines. The thermal mass values are shaded in blue (left) and yellow (right) for reference, where an indicative uncertainty of 5\%, mainly from relativistic effects, is shown.
\medskip}
\end{figure}

The dominant annihilation channel of the DM triplet and fiveplet, at tree-level, is into pairs of $W^\pm$ bosons, and annihilations into $\gamma$ and $Z$ arise at one loop.\footnote{Annihilations into SM fermions are extremely suppressed by the fact that the initial DM-DM pair is in a spin 0 state, due to the Majorana nature of DM.}.
Annihilation cross-sections into gauge bosons are enhanced, for non-relativistic relative velocities $v$ and for masses $M_{\rm DM} \gg m_W$, by the so-called Sommerfeld enhancement~\cite{Hisano:2004ds}.
This effect grows for smaller velocities, and saturates for $v \lesssim 10^{-2} c$, so that both the Milky Way and the dSph's belong to the ``saturated'' regime.
The resulting values of the velocity-weighted annihilation cross section $\langle \sigma v \rangle$ (hereafter called annihilation cross section) are shown in fig.~\ref{fig:NR_Sommerfeld_XS}, for all the possible gauge boson final states, and are computed along the lines of~\cite{Cirelli:2015bda}.
The resonant structure is reminiscent of the presence of bound states of the DM-DM system, due to EW interactions. Indeed the exchange of EW gauge bosons leads, in the non-relativistic limit, to an effective attractive (or repulsive, for the dips \cite{Chun:2012yt}) potential, which can enhance (or suppress) the tree-level $\langle \sigma v \rangle$ by orders of magnitude.
Notice also the relatively large values of the annihilation cross-section into $\gamma\gamma$ and in $Z\gamma$, responsible for the signal  that is the subject of this paper. We refer the reader to~\cite{Beneke:2014gja,Baumgart:2015bpa} for more details about state-of-the-art computations of the Sommerfeld enhancement, and to~\cite{ArkaniHamed:2008qn} for a very clear explanation of the effect.

\section{Gamma Ray Lines Searches}
\label{sec:gammas}

\subsection{Lines accompanied by a continuum}
\label{sec:lines_continuum}
Annihilating DM particles produce $\gamma$-rays via prompt emission and secondary radiation. The former refers to all the photons produced at the particle physics level, the latter to the photons coming from interactions, with the galactic medium, of other DM annihilation products. (The main processes giving rise to secondary photons are inverse Compton, bremsstrahlung and synchrotron radiation off $e^\pm$.)

Concerning the photon energy spectra, they can be roughly classified according to their shape as follows
\begin{itemize}
\item[$\circ$] $\gamma$-ray lines, from the direct annihilation of two DM particles into $\gamma\gamma$ or $\gamma X$, \textit{e.g.} with $X = Z$. In this case, the spectrum develops a prominent line centered at $E_\gamma = M_{\rm DM}$.\footnote{$\gamma$-ray spectral features other than lines, like for example ``boxes'' or ``triangles'', can also arise in particular models~\cite{Ibarra:2012dw,Ibarra:2016fco}, but we will not treat them here.}

\item[$\circ$] $\gamma$-ray continuum, a smooth spectrum peaked at energies smaller than $M_{\rm DM}$ and, depending on its origin, very broad. Such spectra originates both from prompt emission (from the showering, hadronization and decays of the channels in which DM directly annihilates), and from secondary radiation.
\end{itemize}

The focus of this work is on the first kind of spectral shape, $\gamma$-ray lines.
However, a $\gamma$-ray line signal at a given energy would be always accompanied by a broader and lower energy spectrum of photons, because DM generically annihilates to other channels, giving rise to a $\gamma$-ray continuum as discussed above.
Different relative sizes of the DM annihilation channels lead to different lower energy photon spectra which are, in principle, distinguishable.
Supposing a signal in $\gamma$-ray lines will be observed, then the size of the continuum with respect to the line and its spread in energy could, in principle (if the relative background can be understood), allow to distinguish between  different models in explaining the origin of the line. Also, performing a search for a line plus continuum in photons could allow us to set stronger constraints on a particular model, with respect to a search for $\gamma$-ray lines only.

Inspired by the last consideration, in section~\ref{sec:results} we will compare the \CTA\ sensitivity to lines-only with respect to that to lines plus continuum.
While the line-only analysis can be applied to a large class of DM models, the line plus continuum case requires  knowledge of the sizes of all the dominant DM annihilation channels, and thus the specification of a model.
We will therefore perform the line plus continuum analysis only for the specific benchmarks of sec.~\ref{sec:multiplets}, as an example. In doing so, we will determine the continuum as coming just from the prompt emission, since the secondary radiation for the channel we are interested in (massive EW gauge bosons) is negligible\footnote{Notice that this would not be the case for DM annihilation into $e^+e^-$ and $\mu^+\mu^-$~\cite{Lefranc:2015pza}.
}.

\subsection{Fluxes at Earth}
\label{sec:fluxes}
For the prompt emission, the differential  $\gamma$-ray flux at Earth of self-conjugated DM particles, coming from a given angular direction ${\rm d} \Omega$ of the sky, is writen as
\beq
\label{eq:theoryflux}
\frac{{\rm d} \Phi_\gamma}{{\rm d} \Omega \, {\rm d} E_\gamma}=
\frac 1{8\pi M_{\rm DM}^2}\sum_f \langle \sigma v \rangle_f \frac{{\rm d} N^f}{{\rm d}E_\gamma} \, \frac{{\rm d} J(\theta)}{{\rm d} \Omega} \ ,
\eeq
where $\theta$ is the angle between the direction to the center of a given astrophysical object and the line of sight (l.o.s.).
Here ${\rm d} N^f/{\rm d}E_\gamma$ and $\langle \sigma v \rangle_f$ are the energy spectrum of photons per one annihilation and the annihilation cross-section into a primary channel $f$, respectively.

\begin{figure}[t!]
\centering
\includegraphics[width=0.495\textwidth]{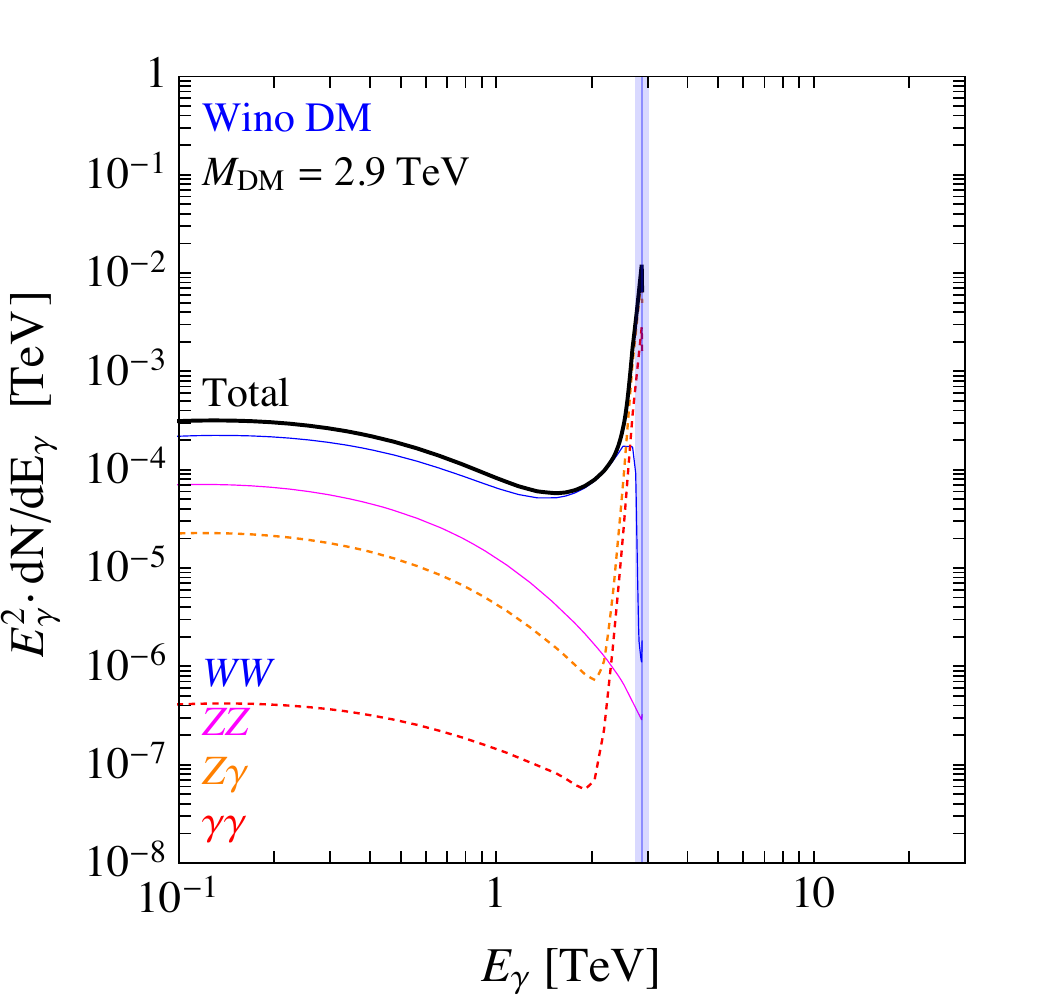}
\includegraphics[width=0.495\textwidth]{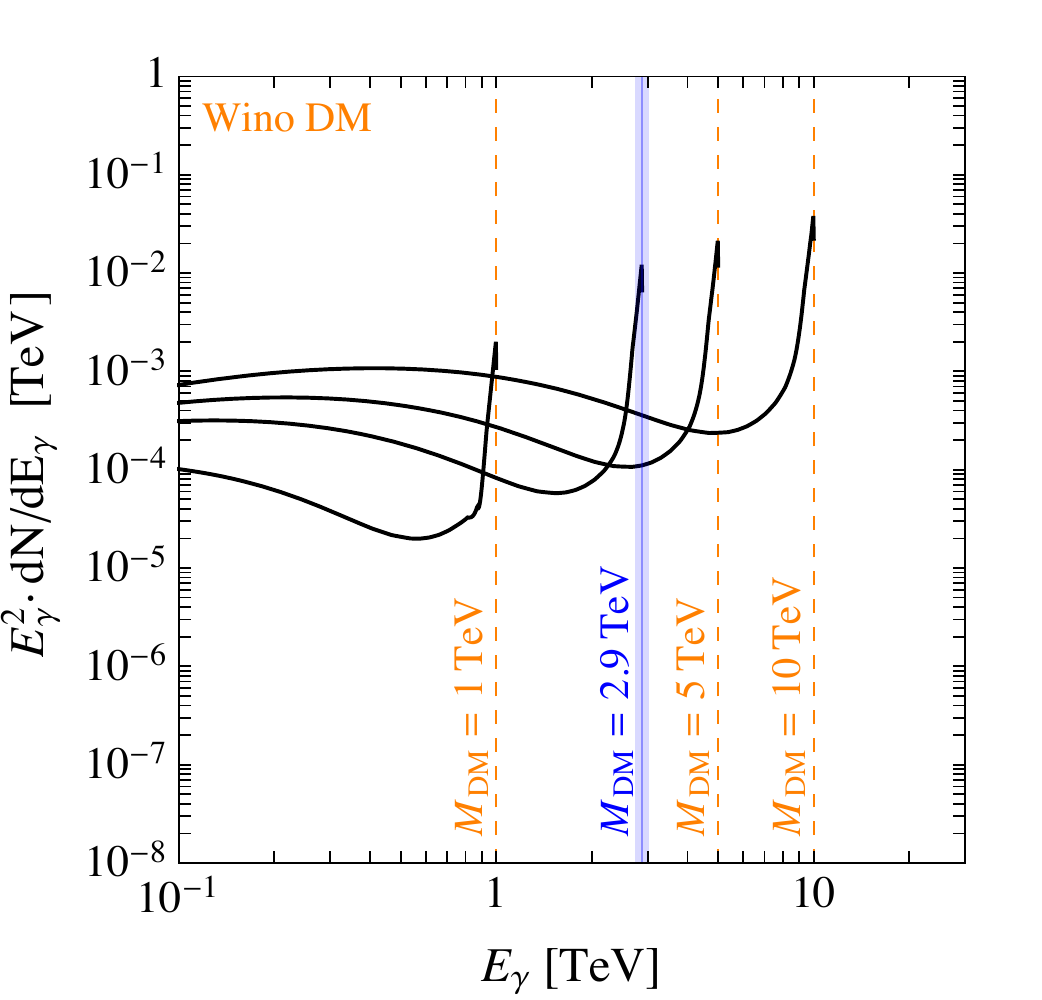} 
\includegraphics[width=0.495\textwidth]{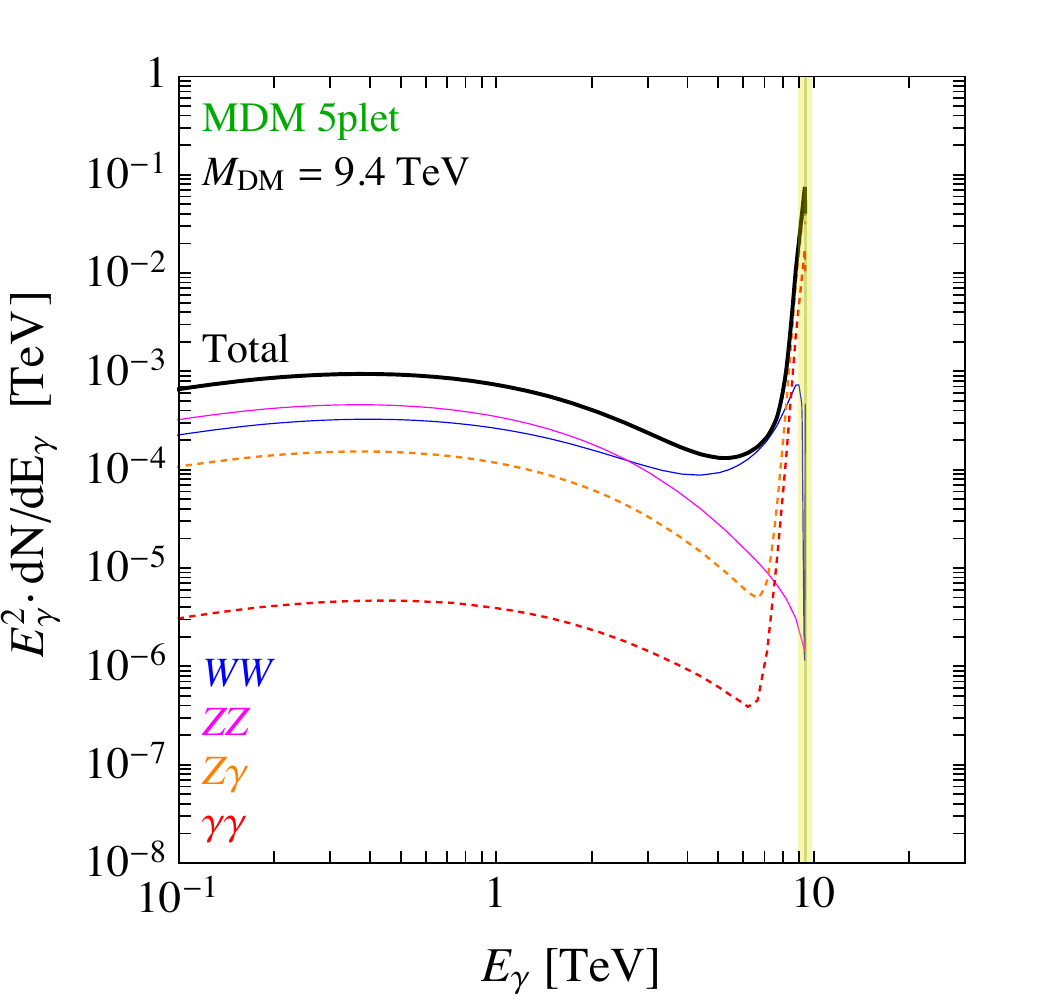}
\includegraphics[width=0.495\textwidth]{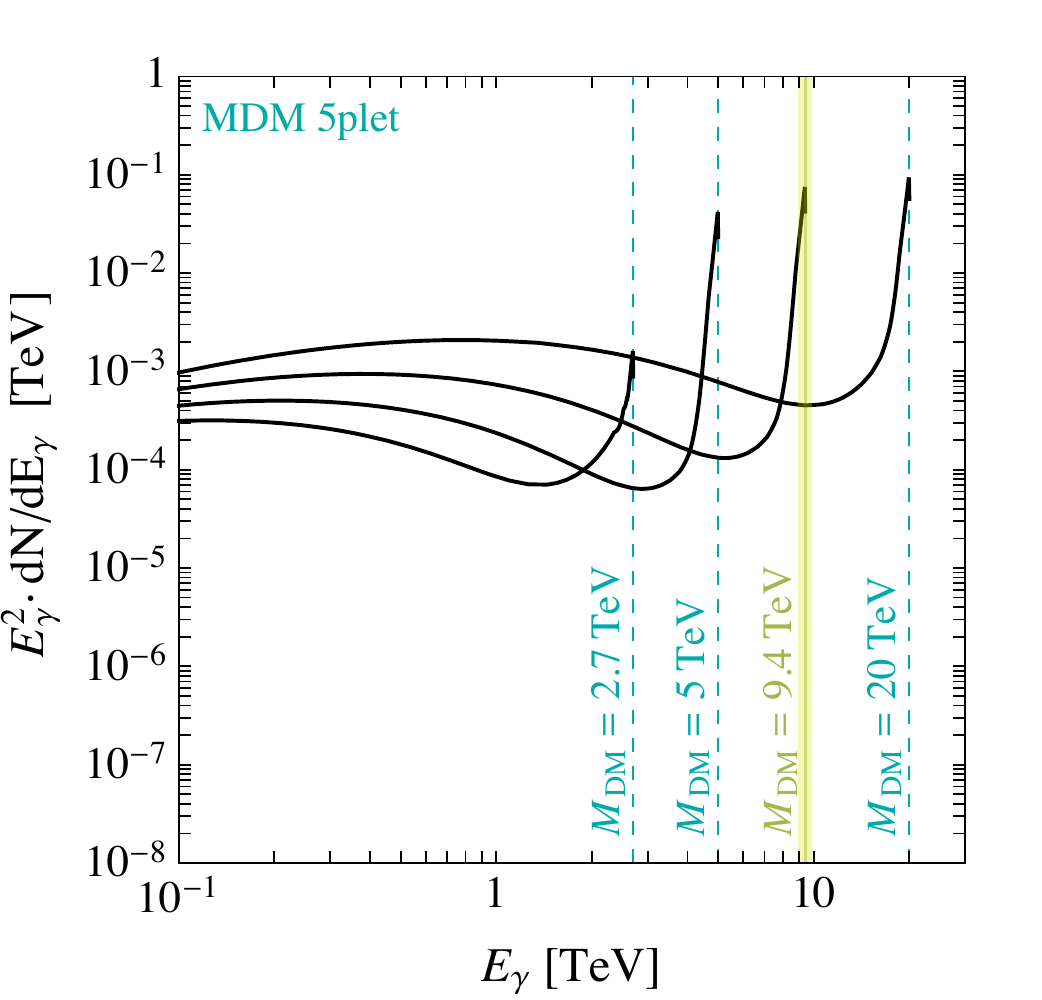} 
\caption{\em \small \label{fig:Normalized_gammaFlux} {\bfseries Prompt Fluxes of Photons.} Total photon flux for a given DM mass shown in black. The different contributions to the total photon fluxes are shown, for the thermal masses, in the left-hand panels (colour coding as in fig.~\ref{fig:NR_Sommerfeld_XS}). The right-hand panels show the total photon fluxes only, for different values of the DM mass. Vertical shadings as in fig.~\ref{fig:NR_Sommerfeld_XS}.
\medskip
}
\end{figure}

When deriving model-independent sensitivities, the spectrum of photons is simply
${\rm d} N^f/{\rm d}E_\gamma = 2 \delta(E_\gamma-M_{\rm DM})$.
When considering the DM candidates of sec.~\ref{sec:multiplets} instead, we compute the DM cross-section in a given annihilation mode $f$ (see fig.~\ref{fig:NR_Sommerfeld_XS}), ${\rm d} N^f/{\rm d}E_\gamma$ by means of the tools provided in~\cite{Cirelli:2010xx}, and we obtain the total photon spectra
\beq
\frac{{\rm d} N}{{\rm d}E_\gamma} =
\frac{1}{\langle \sigma v \rangle_{\rm tot}}\sum_f \langle \sigma v \rangle_f \frac{{\rm d} N^f}{{\rm d}E_\gamma}\ .
\eeq
They are shown with black lines in fig.~\ref{fig:Normalized_gammaFlux}.
In the left panels of fig.~\ref{fig:Normalized_gammaFlux}, we also report the individual contributions to the same spectra, from all the diboson channels. Notice that a line-like feature is given not only by the final states containing a photon, but also by the $W^+W^-$ one. This is because the radiation of a $\gamma$ off a $W^\pm$ is Sudakov-enhanced when the $\gamma$ takes away all the $W^\pm$ energy $M_{\rm DM}$ (apart of course for $m_W$, explaining the behaviour close to $M_{\rm DM}$ of the blue lines), see~\cite{Ciafaloni:2010ti} for more details. The difference of this feature from a pure $\gamma$ line (like the red and orange ones) is however within the energy resolution of both \HESS\ and \CTA.
In the right panels of fig.~\ref{fig:Normalized_gammaFlux}, for reference, we provide the total photon spectra also for mass values other than the thermal ones, \textit{i.e.} 1, 5 and 10 TeV for the Wino, and 2.7, 5 and 20 TeV for the 5plet. We choose the first two mass values as examples of different ratios of the cross-section in $W^+W^-$ with respect to other channels (see fig.~\ref{fig:NR_Sommerfeld_XS}), from which the different ratios between the heights of the peak and that of the lower energy continuum, in the right panels of fig.~\ref{fig:Normalized_gammaFlux}.

In eq.~(\ref{eq:theoryflux}) all the information about astrophysics is included in the so called {\it $J$-factor}, that sets the normalisation of the $\gamma$-rays fluxes at Earth from different targets. The $J$-factor is the l.o.s. integral of the DM energy density square,
\beq
\frac{{\rm d} J(\theta)}{{\rm d} \Omega} = \int_{\rm l.o.s} \rho_{\rm DM}^2\left[r(s,\theta)\right] {\rm d} s \ ,
\label{eq:Jfactors}
\eeq
where $s$ is the l.o.s. coordinate. In the next section, we review the targets used in our analysis  together with the  definition of the regions of interest.

\subsection{Galactic Center}
\label{sec:GC}
Thanks to its vicinity to Earth and to the possibile cusp in the DM density distribution, the Galactic Center is often considered to be  the most promising laboratory for looking at DM annihilation signals.
\paragraph{Uncertainties from baryons.} This widespread assumption is  challenged by the fact that the mass within the innermost kpc of the Milky Way is dominated by baryons.
First, the high density of baryons leads to an expectation of higher astrophysical backgrounds.
Second, and perhaps more important for this work, the baryon dominance over DM implies that astronomical observations are far from probing the DM component of the GC  (see \textit{e.g.}~\cite{Iocco:2015xga} but also see \cite{Wegg:2016jxe} for an alternative view), 
and that numerical simulations need to include hydrodynamics and feedback physics in addition to the gravitational effects, resulting in uncertainties in the obtained DM profiles towards the GC.
More precisely, while most hydrodynamical simulations~\cite{DiCintio:2013qxa,Marinacci:2013mha,Tollet:2015gqa} agree that the DM profile of MW-like galaxies develops a peak towards the GC, their resolution prevents them from making any statement about the profile for radii smaller than one to a few kpc, a region where the impact of baryons is likely to be even more important.
The community has not yet reached an agreement on the best way to model the baryon physics close to the GC, and indeed results  in contrast to those previously  cited have also been found, such as  cores of MW-like galaxies extending to $\sim 5$~kpc~\cite{Mollitor:2014ara}.
\paragraph{DM profiles in this study.}
As long as a clearer determination of the DM profile in the inner MW regions is not achieved, it is worth considering a range of possible choices. Therefore, in our analysis, for the DM density distribution $\rho_{\rm DM}$, we consider the cases of Einasto (peaked), Burkert (cored)\footnote{While the choice of a Burkert  profile appears to be in contrast with most numerical simulations, it has been argued~\cite{Nesti:2013uwa} to be favoured by observations.
}
and NFW with a core $r_{\rm c}$ that we are free to vary. 
For the Einasto and Burkert profiles, the functional forms read
\begin{eqnarray}
\rho_{\rm Ein}(r) &=& \rho_{\rm s} \, \exp\left(-\frac{2}{\alpha} \, \left[\left(r/r_{\rm s}\right)^\alpha-1\right]\right) \ , \label{eq:Einasto} \\
\rho_{\rm Bur}(r) &=&   \frac{\rho_{\rm s}}{\left(1+r/r_{\rm s}\right) \left(1+\left(r/r_{\rm s}\right)^2 \right)} \ , \label{eq:Burkert}
\end{eqnarray}
where $\alpha=0.17$. The functional form of the cored NFW is
\beq
\rho_{\rm NFW}(r, r_{\rm c}) =  
\left\{\begin{array}{l}
\displaystyle \rho_{\rm s} \, \frac{r_{\rm s}}{r_c} \frac1{\left(1+\left(r_c/r_{\rm s}\right)^2\right)} \qquad \mbox{for } r \leq r_{\rm c} \\
\displaystyle \rho_{\rm s} \, \frac{r_{\rm s}}r \frac1{\left(1+\left(r/r_{\rm s}\right)^2\right)} \qquad \mbox{for } r  > r_{\rm c} 
\end{array}\right. \ .
\label{eq:NFWcored}
\eeq
Once $r_c$ is fixed, the above functions depend on two free parameters, the typical scale density $\rho_{\rm s}$ and the scale radius $r_{\rm s}$.
One would ideally choose values that match observations of DM properties in the MW, such as the DM density at the Sun position $\rho_{\odot}$ and the total DM mass within a certain radius. However, precise agreement on the above quantities is still lacking, for recent studies see \textit{e.g.}~\cite{Pato:2015dua,Pato:2015jbx,Huang:2016} for $\rho_{\odot}$, and~\cite{2013ApJ...768..140B,Gibbons:2014ewa,Wang:2015ala,Huang:2016} for the DM mass within a given radius.
%
%
%
The purpose of this study is not to quantify the impact of the latter uncertainties, so we remain for definiteness with the procedure of~\cite{Cirelli:2010xx}, fitting the profiles of eqs.~(\ref{eq:Einasto}), (\ref{eq:Burkert}), (\ref{eq:NFWcored}) to
\beq
\rho(r_{\odot} = 8.33~{\rm kpc}) = 0.3~{\rm GeV/cm}^3 ,\qquad M(< 60~{\rm kpc}) = 4.7\times 10^{11}~M_\odot\,,
\label{eq:fitted_values}
\eeq
where the values in eq.~(\ref{eq:fitted_values}) are well within the uncertainties of the observations cited above.

\paragraph{$J$-factors.}
We first specify an angular region of observation, centred on the GC, that we define as
\beq
\Omega_{\rm obs} = \Omega(\theta < 1^\circ, |b|>0.3^\circ),
\label{eq:GCregion}
\eeq
where $\theta>0^\circ$ is the polar angle with respect to the GC direction, and $b$ is the latitude with respect to the galactic disk plane. For convenience of the reader, we also give the corresponding distances from the GC, $1^\circ \simeq 145$ pc and $0.3^\circ \simeq 44$ pc for $r_{\odot} = 8.33$ kpc.
The region of interest $\Omega_{\rm obs}$ is the same one chosen by \HESS\ for its GC observations~\cite{Abramowski:2013ax,Lefranc:2015vza}, and is motivated by the need to mask the large photon backgrounds from the galactic ridge $|b|<0.3^\circ$, see sec.~\ref{sec:backgrounds}.

We can now evaluate the $J$-factor of eq.~(\ref{eq:Jfactors}), integrated over $\Omega_{\rm obs}$, necessary to compute the total DM annihilation signal from the GC via eq.~(\ref{eq:theoryflux}).
The integrated $J$-factors we obtain are given in table~\ref{tab:Jfactors}, where for the cored NFW of eq.~(\ref{eq:NFWcored}) we choose three values of $r_c$ for illustrative purposes.
\begin{table}[tb!]
\renewcommand{\arraystretch}{1.4}
\centering
\small
\begin{tabular}{r|c|c|ccc}
  &  Einasto  &  Burkert & NFW$_{30 {\rm pc}}$ & NFW$_{300 {\rm pc}}$ & NFW$_{3 {\rm kpc}}$\\
\hline
\hline 
 \vspace{.1 cm}

log$_{10} \left(J_{\Omega_{\rm obs}} \left[\frac{\text{GeV}^2}{\text{cm}^5}\right]\right)$ & 21.15 & 18.93 & 21.04 & 20.63 & 19.38 
\end{tabular}
\caption{GC integrated $J$-factors, with NFW core radii reported as pedices.}
\label{tab:Jfactors}
\end{table}

\subsection{Dwarf Spheroidal galaxies}
\label{sec:dSph}

\begin{table}[tb!]
\centering
\small
\renewcommand{\arraystretch}{1.4}
\begin{tabular}{c|c|c|c}
  &  Hemisphere  &  Zenith angle $\theta_z$ & log$_{10} \left(J_{0.5^\circ} \left[\text{GeV}^2/\text{cm}^5\right]\right)$  \\
\hline
\hline 

\Dra & N &  $18^\circ$ &  $19.09^{+0.39}_{-0.36}$ \\
\hline
\Tri & N & $25^\circ$ & $20.25^{+1.28}_{-1.56}$  \\
\hline
\Scu & S &  $25^\circ$ &  $18.43^{+0.38}_{-0.17}$ \\
\end{tabular}
\caption{Integrated $J$-factor together with its statistical errors for each dSph. We also report the celestial location and the mean zenith angle with respect to the location of the \CTA\ sites, using~\cite{dwarf:finder}.}
\label{tab:Jfactors_dSphs}
\end{table}

Dwarf Spheroidal galaxies of the Milky Way  are very clean laboratories to look for DM in high energy $\gamma$-rays. They are in fact one of the most DM dominated objects known in the Universe (see \cite{Courteau:2013cjm} and references therein), their baryonic content being very small both in terms of recent stellar formation activity \cite{Weisz:2011gp, Brown:2012uq} and of interstellar gas medium \cite{Mateo:1998wg}.
After a period of relatively low activity, over which only 9 luminous ``classical'' dSph galaxies were identified as Milky Way satellites, the field of dSph galaxies is now experiencing a rejuvenation thanks to the Sloan Digital Sky Survey (SDSS)~\cite{Abazajian:2008wr}, its successor (the Sloan Extension for Galactic Understanding and Evolution (SEGUE)~\cite{Yanny:2009kg}) and to forthcoming imaging surveys such as the Panoramic Survey Telescope and Rapid Response System (Pan-STARRS)~\cite{PanSTARRS,Kaiser:2002zz}, the Dark Energy Survey (DES)~\cite{DES,Flaugher:2004vg} and the  forthcoming Large Synoptic Survey Telescope (LSST)~\cite{LSST,Tyson:2002nh,Hargis:2014kaa}.
The rate of discovery of faint surface-brightness objects, usually called ``ultra-faint'' dSphs,  has been dramatically increasing, making available a relatively large sample of new targets for DM indirect detection. 

\paragraph{Uncertainties in the $J$-factor determination.}
Concerning indirect DM detection, the integral of the astrophysical factor in eq.~\eqref{eq:Jfactors} is determined by constructing dynamical mass models based on the Jeans equation.
Several studies have estimated the $J$-factor in many dSphs along with its statistical error~\cite{Strigari:2007at, Martinez:2013els, Geringer-Sameth:2014yza, Bonnivard:2015xpq}, and several others have used it to constrain the annihilation cross-section of DM~\cite{Evans:2003sc, Essig:2009jx, GeringerSameth:2011iw, Charbonnier:2011ft, Abramowski:2013ax, Ackermann:2015lka, Aleksic:2013xea, Ackermann:2015zua}. In particular, once a specific approach is adopted, the high quality of kinematical data is in general sufficient to provide small statistical errors\footnote{This is particularly true in the case of ``classical'' dSphs, which are characterized by a large number of stellar tracers.
}.
Nevertheless, there are crucial systematic uncertainties that come with the determination of the $J$-factor, that are not taken into account in most of these studies (see e.g.~Sec.~3.4 of~\cite{Lefranc:2016dgx}, where a thorough description of the sources of systematic uncertainties is provided).
These systematic errors of the theoretical modelling can have a large impact on the  determination of the parameters of a given dSph, including the evaluation of the $J$-factor, especially for candidates with a small number of stellar tracers.  Therefore it is not surprising that, in the past few months, several papers have tried to address the impact of some of these systematics~\cite{Hayashi:2016kcy, Ullio:2016kvy, Evans:2016xwx, Genina:2016kzg}.

\paragraph{Choice of the dSphs and $J$-factors.}
We base our analysis on~\cite{Hayashi:2016kcy}  which evaluates the $J$-factors taking into account the systematic errors coming from non-spherical dark halos, by using axisymmetric mass models.
We use this work for two main reasons: $i)$ they use the most recent kinematic data of the  most complete set of dSphs (7 ``classical'' and 17  ``ultra-faint''); $ii)$ they adopt, for the first time, an  unbinned analysis for comparison between  models and data.
This is a more robust and conservative method for constraining dark halo parameters, especially in faint dSph galaxies where the number of stellar tracers is very small. 

From their list of dSph galaxies, we choose those with the best $J$-factors. In the northern hemisphere we take  the  classic dSph \Dra\ and the ultra-faint dSph \Tri.  In the southern hemisphere we consider the classical dSph \Scu. In Tab.~\ref{tab:Jfactors_dSphs} we provide, for each dSph, the logarithmic values of the $J$-factor (integrated over $0.5^\circ$ angular annulii surrounding each dSph) together with its statistical error. We also report in the second and third columns the celestial location and zenith angle $\theta_z$ respectively.
Notice that, while \Dra\ and \Scu\ have hundreds of tracers, the DM properties of \Tri\ derived in~\cite{Hayashi:2016kcy} are based only on 13 tracers, making the candidate more speculative cause potentially more subject to systematics.

\section{CTA sensitivities}
\label{sec:CTA}

\CTA\ is the forthcoming observatory for $\gamma$-ray astronomy and, in a possible design scenario, it is envisaged to be a two-site array. The northern hemisphere array of \CTA, located in Roque de los Muchachos Observatory in La Palma in the Canary Islands, will consist of four large-size (23m diameter) telescopes and 15 medium-size (12m diameter) telescopes. The southern hemisphere array of \CTA, located in the European Southern Observatory Cerro Paranal in Chile, would consist of 4 large-size, 24 medium-size and 72 small-size telescopes, in order to cover the full energy range. 

\paragraph{CTA performances.} \CTA\ will provide excellent angular resolution and very wide energy range with respect to  any operating Cherenkov telescope instrument. Although the final configuration of the array is not yet decided, performance studies based on  Monte Carlo (MC) simulations have been performed on many candidate arrays, in order to figure out the response functions in terms of flux sensitivity, energy and angular resolutions and background rejection.
In the present article, we use the array layout for the southern and northern sites of the \CTA\ observatory as given in~\cite{CTA:newperformances} (which supersedes previous studies, like~\cite{Bernlohr:2012we}, performed when the \CTA\ sites had not been fixed yet).

\subsection{DM signals and irreducible backgrounds}
\label{sec:backgrounds}
\begin{figure}[!t]
\centering
\includegraphics[width=0.95\textwidth]{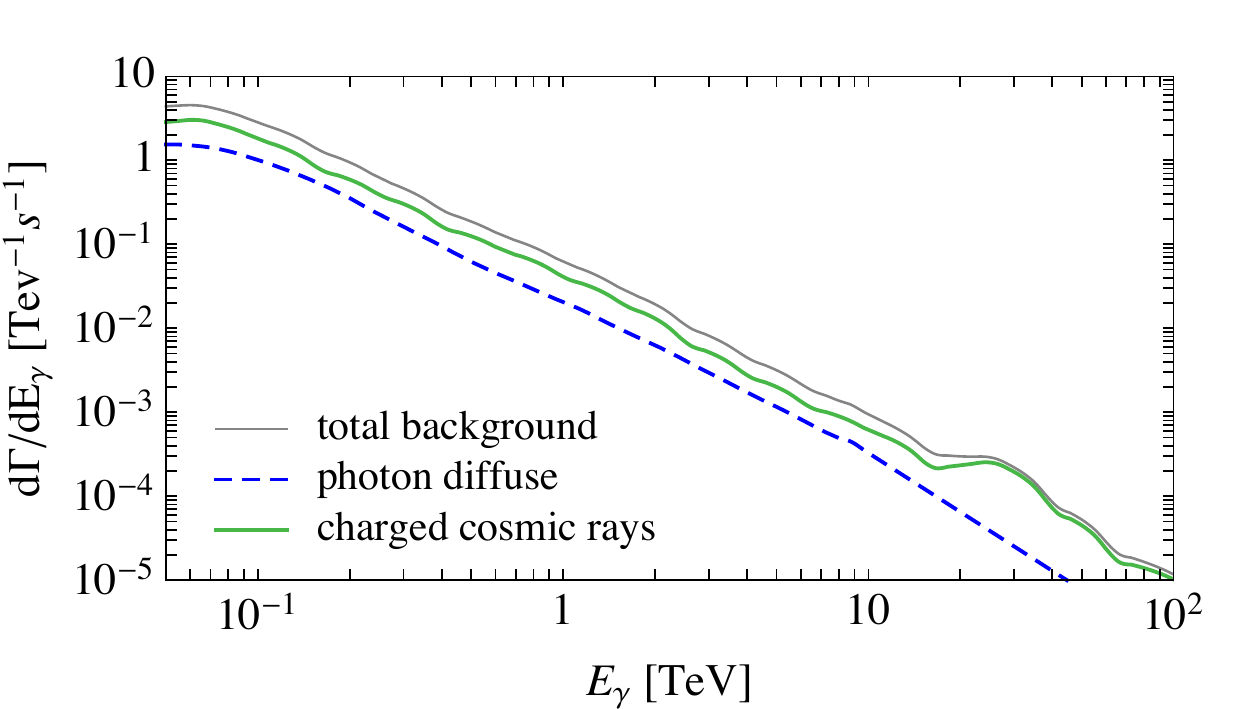}
\caption{\em \label{fig:bkg} \small {\bfseries CTA backgrounds.} Spectral morphology of mean expected count rates of irreducible backgrounds to searches for $\gamma$-ray lines, for the GC region of interest $\Omega_{\rm obs}$ (see text). Green: background from charged CRs, as modelled by \CTA. Blue dashed: diffuse photon emission, from Fermi observations and their extrapolations. Gray thin: sum of the two components.
\medskip}
\end{figure}

The number of photon counts in a given energy bin $\Delta E_i$, for both the signal and the background, is given by
\beq
N_\gamma^i = {\rm observation~time} \times \int_{\Delta E_i} \frac{{\rm d}\Gamma}{{\rm d}E'}(E',\theta_z) {\rm d} E',
\eeq
where we express the counting rate as
\beq
\frac{{\rm d}\Gamma}{{\rm d}E'}(E',\theta_z) =
\int {\rm d} E_\gamma \mathcal{A}_{\rm eff}(E_\gamma,\theta_z) \frac{{\rm d} \Phi_\gamma}{{\rm d} E_\gamma}(E_\gamma) \, \mathcal{R}(E_\gamma,E')\ .
\label{eq:counting_rate}
\eeq
Let us now discuss the different pieces entering the counting rate eq.~(\ref{eq:counting_rate}).
\begin{itemize}

\item[$\diamond$] $\mathcal{R}$ is the energy resolution of the experiment, which we model as a gaussian with variance $\sigma^2$,
\beq
\mathcal{R}(E_\gamma,E') =
\frac{1}{2\pi \sigma^2(E_\gamma)}\exp{\left(-\frac{(E_\gamma-E')^2}{2\,\sigma^2(E_\gamma)}\right)}, \qquad
\sigma(E_\gamma) = \frac{\delta_{\rm res}(E_\gamma)}{\sqrt{8 \ln(2)}}
\eeq
where we take $\delta_{\rm res}$ for the northern and southern sites from~\cite{CTA:newperformances}.\\

\item[$\diamond$]
$\mathcal{A}_{\rm eff}$ is the effective area of the telescope array, where on top of the dependence on the photon energy, we include (for the first time in \CTA\ studies) that on the zenith angle $\theta_z$, because we are interested in targets with different positions in the sky.
A dependence on $\theta_z$ is expected because different zenith angles correspond, for the shower, to different thicknesses of atmosphere to cross and to different projections on the telescopes.
A $\theta_z$ dependence is indeed taken into account in \HESS, but we are not aware of \CTA\ giving this information.
Therefore we assume the effective area of \CTA\ will depend on $\theta_z$ in the same way as the \HESSII\ area\footnote{
We choose the $\theta_z$ dependence of \HESSII, as opposed to the one of \HESS~\cite{Aharonian:2006pe}, because of the \HESSII\ sensitivity to lower energy photons, which makes it a closer analogue to \CTA.
}, and we extract the \HESSII\ dependence from~\cite{Chretien:2016} (see Chapter 4). 
The \CTA\ effective area then reads
\beq
\label{eq:theta_z}
\mathcal{A}_{\rm eff} (E_\gamma,\theta_z) =
\mathcal{A}_{\rm MC}^{\rm CTA} (E_\gamma,0) \times \frac{\mathcal{A}^{\rm HESS-II} (E_\gamma,\theta_z)}{\mathcal{A}^{\rm HESS-II} (E_\gamma,0)}
\eeq
where $\mathcal{A}_{\rm MC}^{\rm CTA} (E_\gamma,0)$ is determined by a \CTA\ MC simulation~\cite{CTA:newperformances}.

\item[$\diamond$] Finally, ${\rm d} \Phi_\gamma/{\rm d} E_\gamma$ is the differential photon flux at Earth:
$i)$ for the signal, it is given by the integration of eq.~(\ref{eq:theoryflux}) over the angular regions of interest (see sec.~\ref{sec:GC} for the GC and sec.~\ref{sec:dSph} for the dSph);
$ii)$ for the background, we include two components, one coming from the misidentified interactions of charged CRs ($p$, $e^\pm$, nuclei) with the atmosphere, and one from the diffuse photon background\footnote{
Its importance for GC $\gamma$-ray searches with \CTA\ was first noted in~\cite{Silverwood:2014yza}, where however it was quantified without any inclusion of systematics. See~\cite{Lefranc:2015pza} for a later treatment.
}. 
Modelling of the former is provided by \CTA\ itself in~\cite{CTA:newperformances}, and we stay with that determination.
To model the latter, we proceed as in~\cite{Lefranc:2015pza} (where the ``RoI1'' coincides with our $\Omega_{\rm obs}$) and extrapolate to higher energies the Fermi measurements of photon Galactic Diffuse Emission (GDE)~\cite{Ackermann:2012pya,Ackermann:2014usa}, which reach a few hundred GeV. To do so, we assume the photon energy spectrum follows the same power law as the Fermi data. 
We then convolve this determination with the \CTA\ sensitivity.
Note that while the misidentified ``charged CR'' component is isotropic, the ``photon diffuse'' one quickly drops away from the GC, and is therefore irrelevant for dSph galaxies.

These two background components are shown in fig.~\ref{fig:bkg} for the GC region (southern \CTA\ site), together with their sum.
We show the backgrounds corresponding to the mean values of the \CTA\ MC simulations, corresponding to the expected means of the sensitivities of sec.~\ref{sec:results}.
One sees in fig.~\ref{fig:bkg} that the inclusion of the photon diffuse component has a non-negligible impact on the total background count.
This is the case because the most recent determination of \CTA\ rejections of charged CRs~\cite{CTA:newperformances}, that we are using, substantially improves with respect to  previous determinations~\cite{Bernlohr:2012we}\footnote{To avoid possible confusion in comparing with the recent~\cite{Ibarra:2015tya}, notice that there the charged CR background is modeled analytically -as opposed to our direct use of \CTA\ results-, and that their photon backgrounds are not relevant for our study, since they come from the inner GC and from the galactic ridge, that we have masked by choosing the region of observation $\Omega_{\rm obs}$ in eq.~(\ref{eq:GCregion}).
}.
We will quantify the impact of the photon diffuse component on the projected sensitivities in sec.~\ref{sec:results}.

\end{itemize}

\subsection{Analysis methodology and likelihood}
\label{sec:likelihood}

To derive the \CTA\ sensitivity, we perform a likelihood ratio statistical test. The total likelihood $\mathcal{L}$ is obtained as 
a product over the energy bins $i$. For a given $m_{\rm DM}$ and astrophysical target considered in our analysis, we have

\begin{equation}
\label{totlik}
\mathcal{L}  (\mDM,\langle \sigma v \rangle) = 
\left\{\begin{array}{l}
\displaystyle \prod_{i}^{\mathcal N}  \mathcal{L}_{i}  (\mDM,\langle \sigma v \rangle)  \ ,\qquad \text{GC} \\
\displaystyle \prod_{i}^{\mathcal N} \underset{J}{\rm Max} \, \left[ \mathcal{L}_{i}  (\mDM,\langle \sigma v \rangle) \times  \mathcal{L}^J(J)\right]  \ ,\qquad \text{dSphs} 
\end{array}\right.
\end{equation}
where the $\mathcal N$ energy bins of the counts spectrum are logarithmically spaced between the lower energy threshold of 30 GeV and 80 TeV.   
Notice that we are not using the spatial information in the likelihood, because the precise spectral location of the line, with respect to the smooth background, provides a discrimination in itself.  This agrees with the procedure of \HESS\ for $\gamma$ lines searches~\cite{Abramowski:2013ax}, as opposed to searches for a continuum spectrum, where the spatial information is crucial~\cite{Abramowski:2011hc,Lefranc:2015pza,Lefranc:2016dgx}.
Let us now discuss the likelihood terms entering in eq.~\ref{totlik}. 
\begin{itemize}

\item[$\diamond$]  Given a set of data $N_{\rm obs}^{i=\{1,... , \mathcal N \}}$,  the individual likelihood $\mathcal{L}_{i} $ in eq.~(\ref{totlik}) is given by a Poisson distribution 
\begin{equation}\label{Likelihood}
\mathcal{L}_{i}  (N_\gamma | N_{\rm obs}) = \frac{\left(N_\gamma^i\right)^{N_{\rm obs}^i}}{N_{\rm obs}^i!}\, e^{-N_\gamma^i} ,
\end{equation}
where $N_\gamma^i $ is the sum of the predicted number of events from DM and those from the irreducible backgrounds discussed in sec.~\ref{sec:backgrounds}. 

\item[$\diamond$] In the case of dSphs, the term $\mathcal{L}^J$ measures the impact of the statistical errors of the $J$-factors. In particular, following~\cite{Ackermann:2011wa} we write
\beq
\mathcal{L}^J (J | J_{\rm obs}, \sigma) = \frac1{\ln(10)  \,  J_{\rm obs} } \, \mathcal G (\log_{10} J | \log_{10}  J_{\rm obs},\sigma) \ ,
\eeq
where  $J_{\rm obs}$ is the observed $J$-factor of a given dSph, $\sigma$ is the uncertainty on $\log_{10} J_{\rm obs}$ (both taken from~\cite{Hayashi:2016kcy}), and $\mathcal G(x|\mu,\sigma)$ is a Gaussian distribution of mean $\mu$ and standard deviation $\sigma$\footnote{
The $J$-factor for the GC is even more uncertain, and we take this into account by showing sensitivities corresponding to much different DM density profiles, see sec.~\ref{sec:GC}.
}.
We then maximise the product in eq.~\ref{totlik} with respect to the nuisance parameter $J$.

\end{itemize}
Our statistical analysis then proceeds adopting a likelihood ratio Test Statistic TS$ = -2 \ln\left[\mathcal{L}(m_{\rm DM}, \langle \sigma v \rangle)/\mathcal{L}_{\rm max}(m_{\rm DM}, \langle \sigma v \rangle)\right]$, where $\mathcal{L}_{\rm max}$ is maximized over the free parameter $\langle \sigma v\rangle$. We then assume that TS follows a $\chi^2$ distribution, so that, for any $m_{\rm DM}$, values of TS higher than 2.71 are excluded at the 95\% confidence level.

To take possible statistical fluctuations of the background  into account, we generate a sample of one thousand background counts for each bin, \textit{i.e.} one thousand sets of the form $N_{\rm bkg}^{i=\{1,... , \mathcal N \}}$ (where $\mathcal N$ is not to be confused with one thousand). To do so, we assume that the number of background events within each bin, $N_{\rm bkg}^i$, vary following a Poissonian distribution. We then perform the likelihood ratio TS for each of the thousand sets generated as above, and we determine the mean expected exclusion as the mean exclusion over this set. We then determine the 68\% and 95\% containment bands from the 680 and 950 closest lines to the mean expected exclusion.

\section{Results and discussion}
\label{sec:results}
In what follows, we always assume 100 hours of observation of the GC\footnote{
The response functions are provided in~\cite{CTA:newperformances} up to 50 hours of observation, to obtain the sensitivities for 100h we simply rescale the 50h ones by $\sqrt{2}$.},
and 50 hours of observation of \Scu, \Dra\ and \Tri. These observation times represent realistic benchmarks, as indeed the GC will arguably be looked at for a longer time than dwarves (keep however in mind the 160 hours of observation of \Seg\ by \MAGIC~\cite{Aleksic:2013xea}).

\begin{figure}[!t]
\centering
\includegraphics[width=0.495\textwidth]{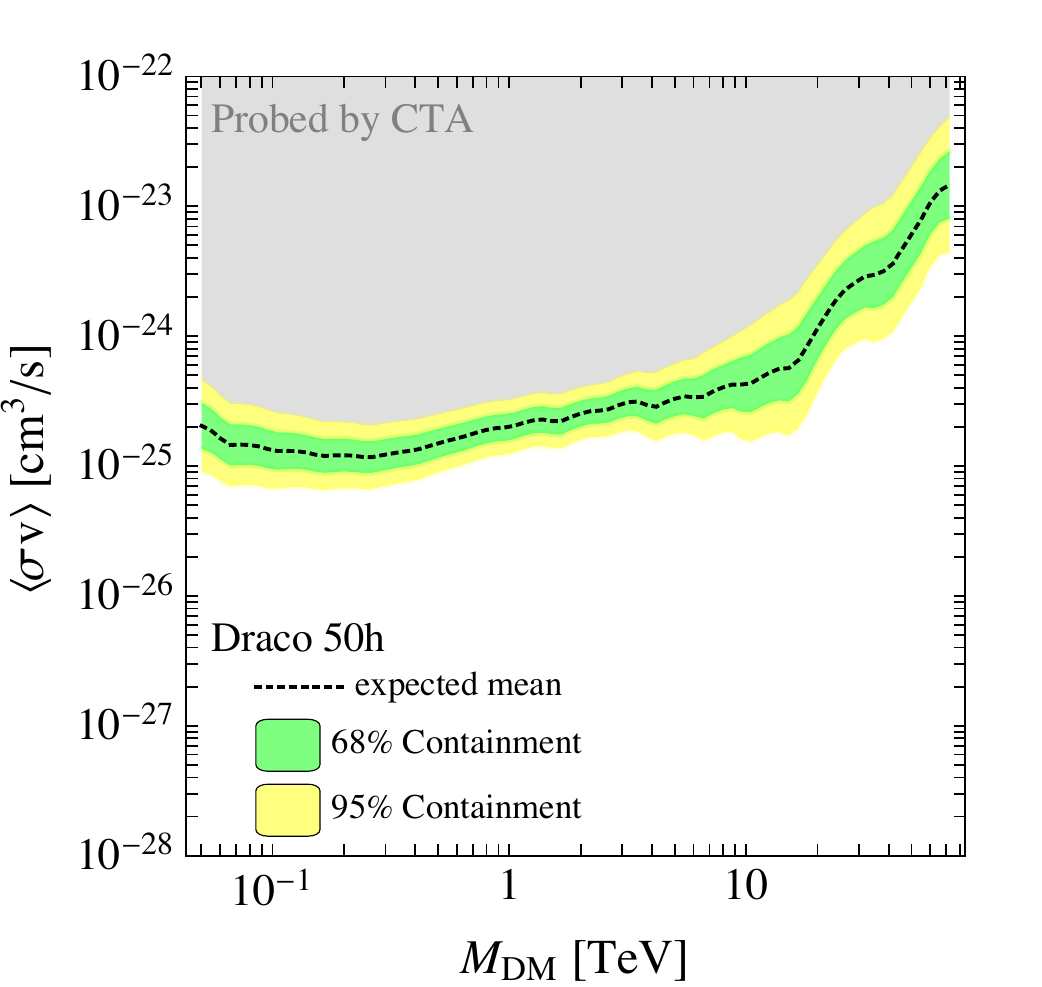}
\includegraphics[width=0.495\textwidth]{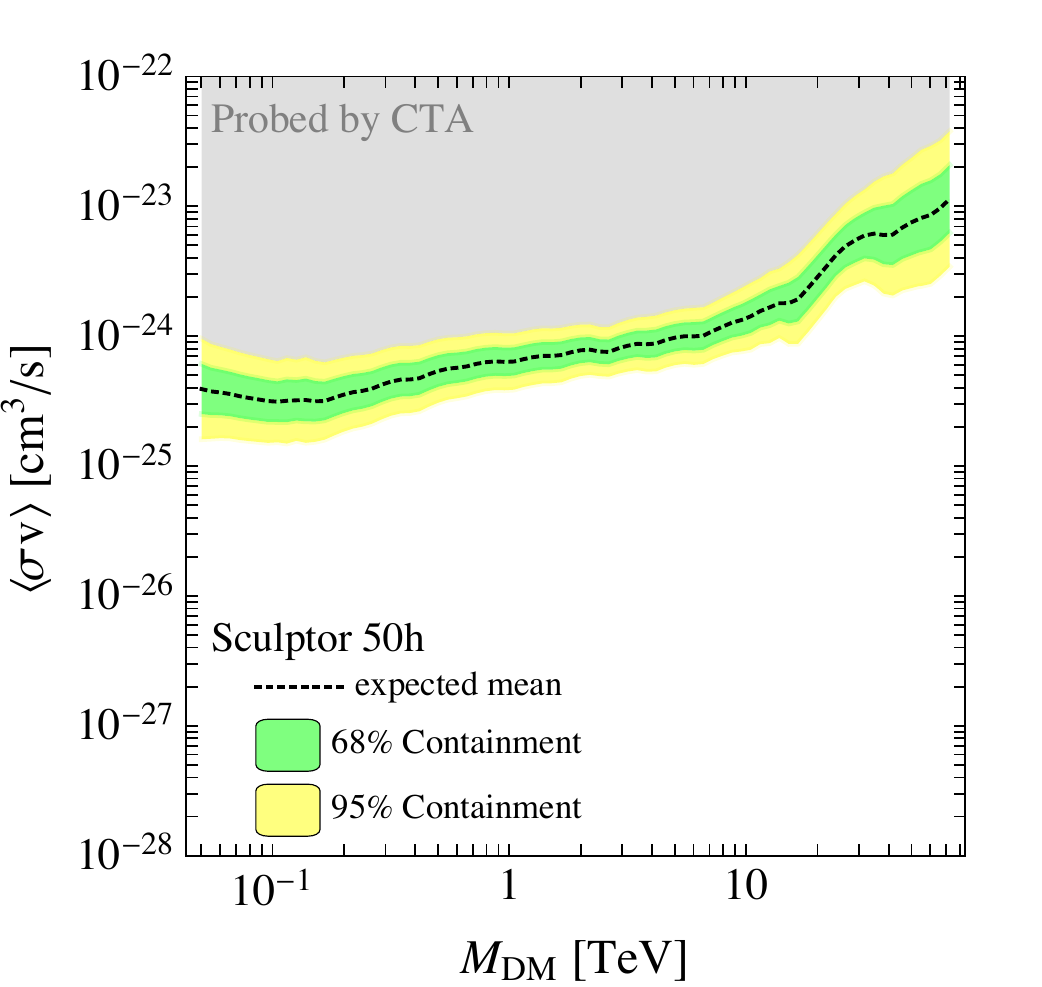} \\
\includegraphics[width=0.495\textwidth]{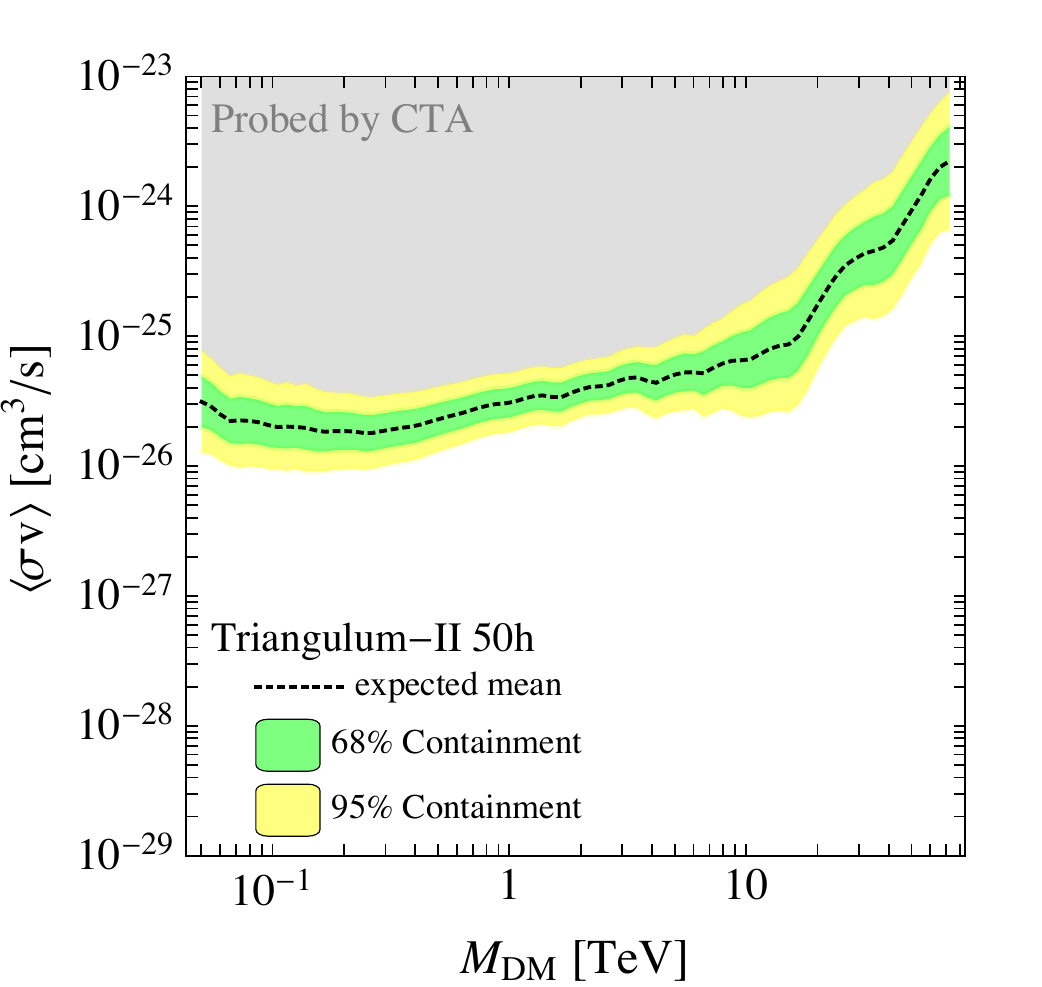} 
\includegraphics[width=0.495\textwidth]{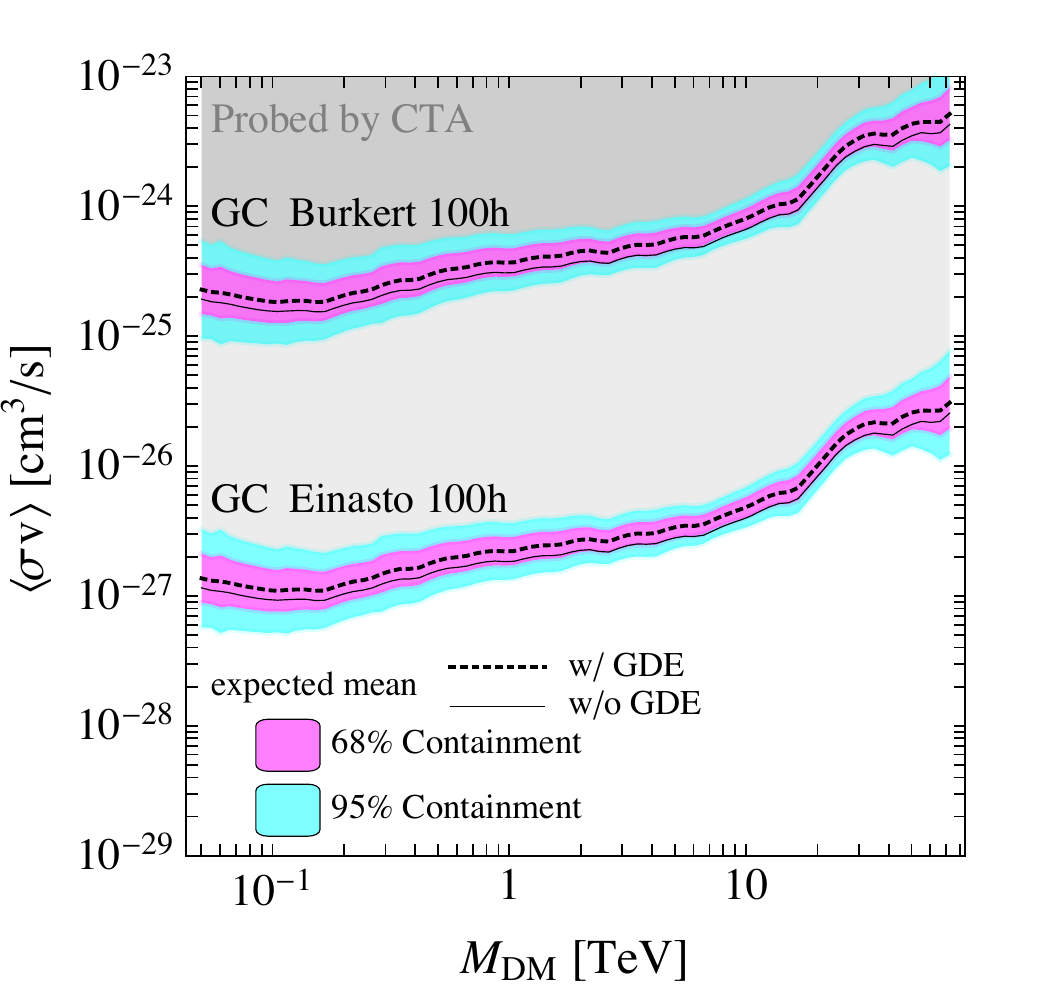} \\
\caption{\em \small \label{fig:model_indep_pessimistic}
\small {\bfseries CTA sensitivities.} Expected mean \CTA\ sensitivities (dashed black lines), with 68\% (green for the dSphs, magenta for the GC) and 95\% (yellow for the dSphs, light blue for the GC) containment bands. For the GC, it is reported also the mean expected sensitivity that one would obtain ignoring the diffuse photon component of the background (thin black line). Left panels: targets observed by the northern site. Right panels: targets observed by the southern site.
\medskip}
\end{figure}
\paragraph{Model independent sensitivitites.} 
The expected \CTA\ sensitivities to $\gamma$-ray line signals are shown in fig.~\ref{fig:model_indep_pessimistic}.
Notice the wider containment bands of the objects observed by the northern site (\Dra\ and \Tri), compared to those observed by the southern site (GC and \Scu). This is a consequence of the better sensitivity of the southern site to high energy $\gamma$-rays, thanks also to the higher number of telescopes, which implies higher acceptance and lower residual background.
It is also worth pointing out that that all the observational targets in this study lie close to each other in observation angle $\theta_z$, so that the differences induced by this dependence (see eq.~\ref{eq:theta_z}) are very mild.
Concerning the GC, on top of the mean exclusion in dashed, we show as a thinner continuous line what the mean exclusion would have been, if we had not included the photon diffuse emission component of the background, see sec.~\ref{sec:backgrounds}.
One sees that the improvement would be within the 68\% containment band, and therefore small compared to statistical uncertainties on the residual background.

\begin{figure}[!t]
\centering
\includegraphics[width=0.495\textwidth]{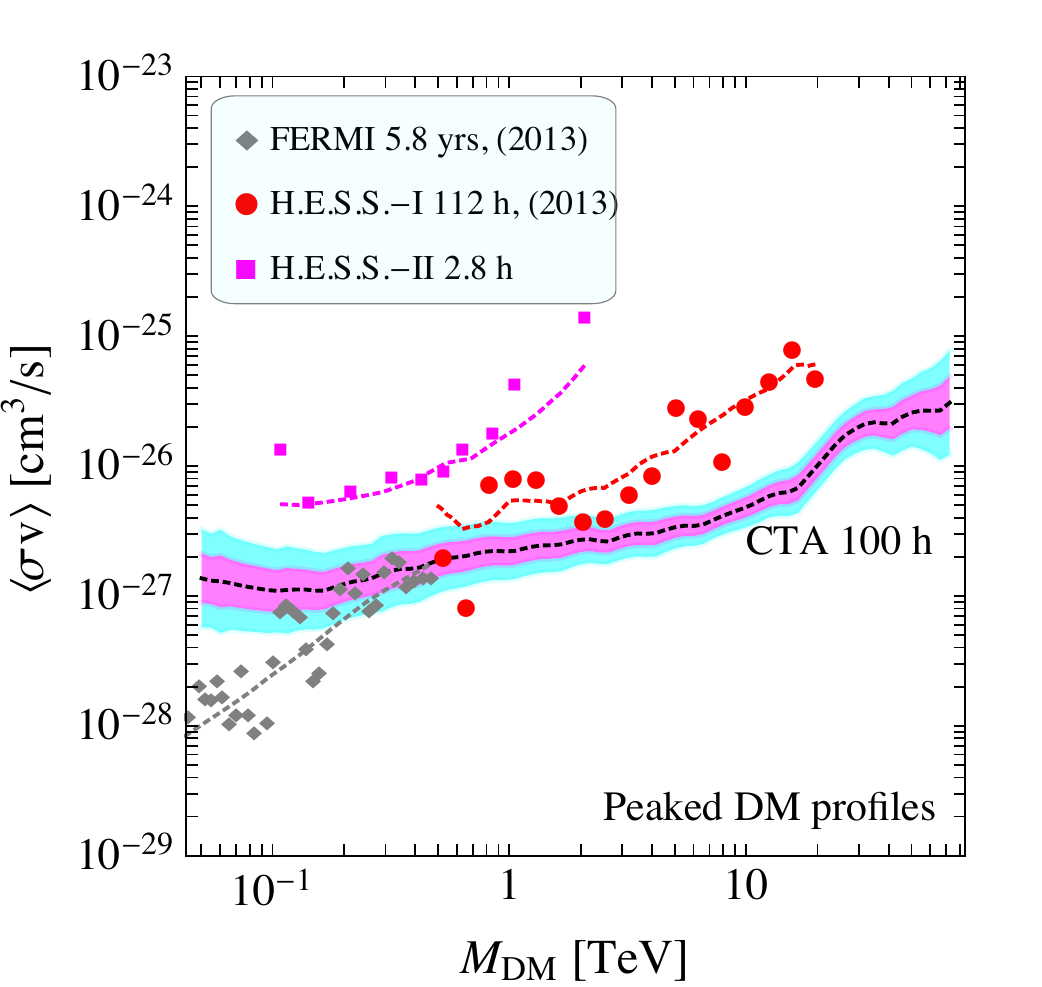} 
\includegraphics[width=0.495\textwidth]{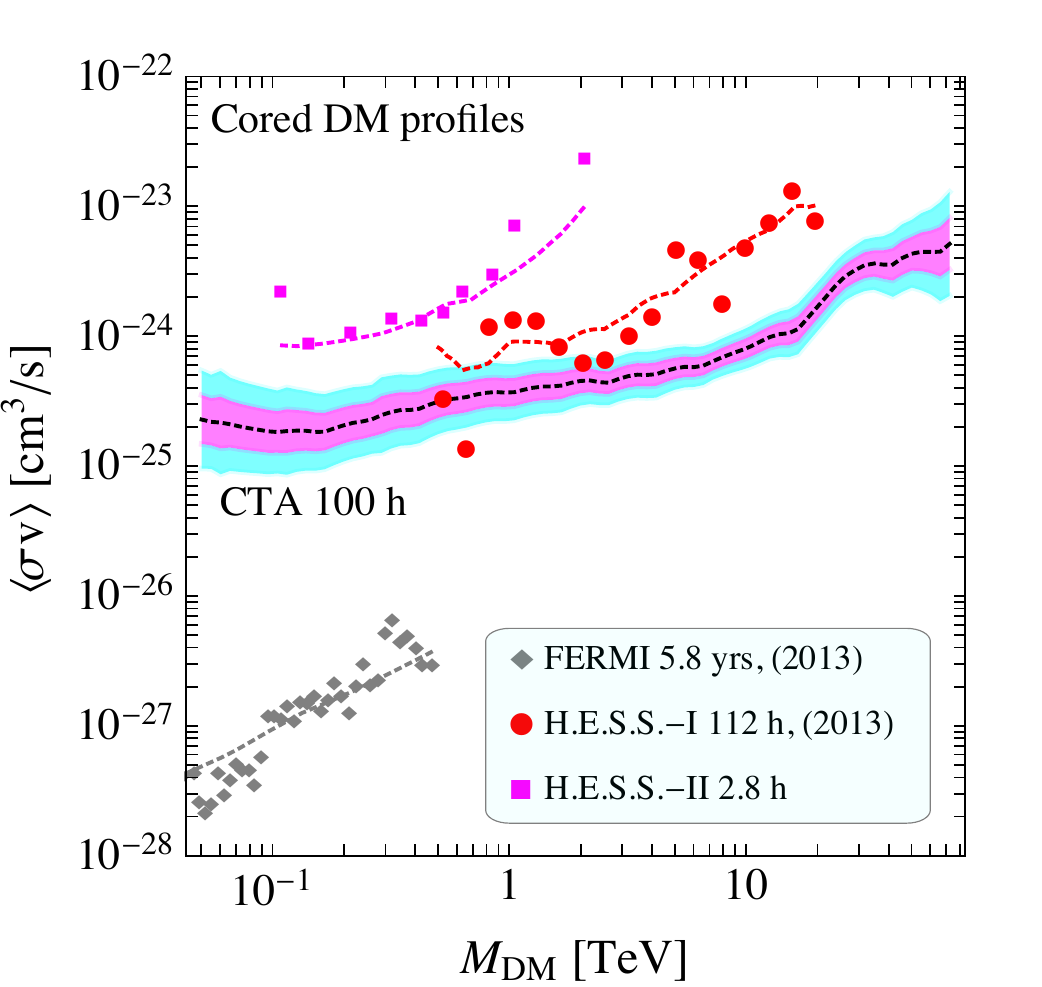}
\caption{\em \small \label{fig:GC_pastfuture} {\bfseries CTA sensitivity vs current exclusions.} Dashed lines: expected mean sensitivities of \CTA\ (100 hours, black), \HESS\ (112 hours, red) and \HESSII\ (2.8 hours, magenta) observations of the GC, and of \FERMI\ (5.8 years, gray) observations of various MW regions (see text). Points: actual exclusions of \HESS\ (circles), \HESSII\ (squares) and \FERMI\ (diamonds), with colour coding and observation times as before. 68\% and 95\% containment bands shown in magenta and light blue for \CTA. Left-hand plot: Einasto profiles for all experiments. Right-hand plot: Burkert profile for \HESS, \HESSII\ and \CTA, Isothermal for \FERMI. 
\medskip}
\end{figure}

\paragraph{Comparison with existing limits.} With respect to presently existing bounds for $\gamma$-ray lines, \CTA\ will constitute a major improvement.
The only existing bound on $\gamma$-ray lines from dSph has been derived by \MAGIC~\cite{Aleksic:2013xea}, from 160 hours of observation of the dSph \Seg. However, the estimate of its $J$-factor has dropped by roughly two orders of magnitudes~\cite{Bonnivard:2015xpq,Hayashi:2016kcy}, compared to the one derived by \MAGIC\ and used to cast their exclusion limits. Therefore, we refrain from showing the \MAGIC\ exclusion on the same figures as our projections.
Concerning the GC, we show in figure~\ref{fig:GC_pastfuture} how our projected \CTA\ sensitivities compare with existing bounds from Fermi~\cite{Ackermann:2015lka}, \HESS~\cite{Abramowski:2013ax} and \HESSII~\cite{Kieffer:2015nsa}.
We show as points the actual limits, and as continuous lines the expected limits corresponding to those points, to guide the eye in the comparison with \CTA. \HESS\ and \HESSII\ observed the same region $\Omega_{\rm obs}$ of our study, and in fig.~\ref{fig:GC_pastfuture} we have rescaled the existing bounds to be consistent with our choices of DM profile (see sec.~\ref{sec:GC}), being $J_{\Omega_{\rm obs}}^{\rm HESS} = 4.4 \times 10^{21}$ GeV$^2$/cm$^5$.
The regions of interest of \FERMI\ change from ours, and we simply report the limits as they give them, for two choices that appear most similar to ours, \textit{i.e.} an Einasto and an Isothermal profiles.
The respective regions of interest, called ``R16" and "R90", are optimized for the related DM profiles, and are much more extended than our $\Omega_{\rm obs}$.
This in particular explains the much better \FERMI\ sensitivity with a cored (Isothermal) profile, which comes from its capability to observe a very large sky region.

\begin{figure}[!t]
\centering
\includegraphics[width=0.495\textwidth]{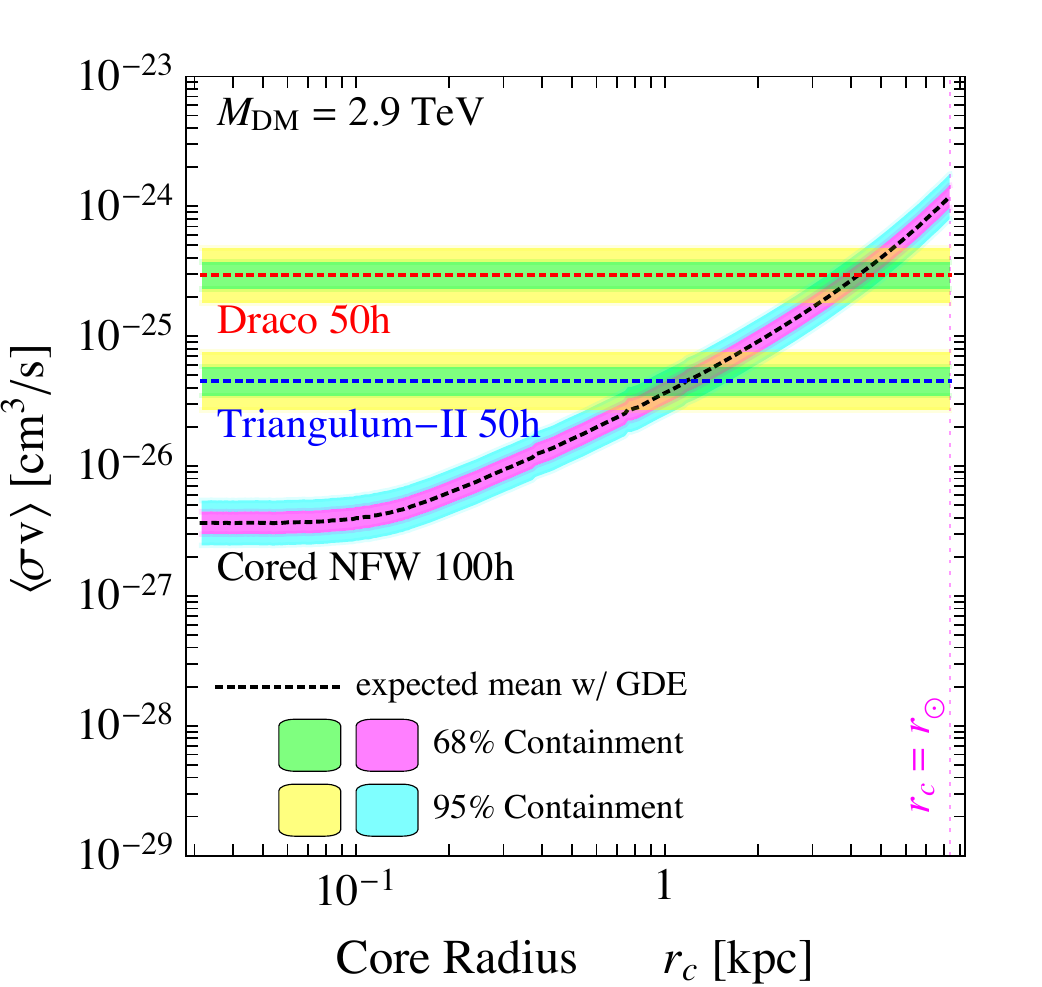} 
\includegraphics[width=0.495\textwidth]{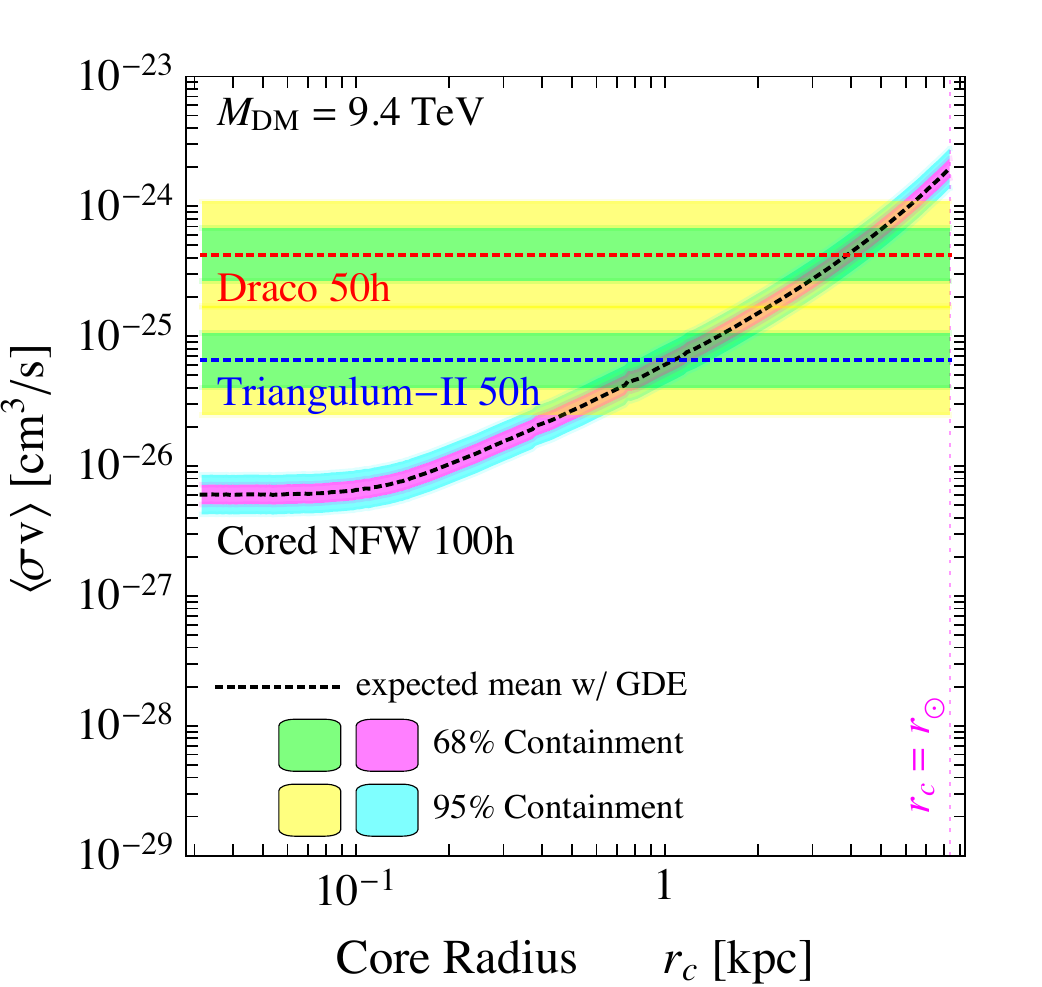} \\
\caption{\em \small \label{fig:GC_vs_dSph} {\bfseries Galactic Center vs dwarf Spheroidals.} \CTA\ sensitivity to monochromatic photons from the GC (100 hours of observation), \Tri\  (50 hours) and \Dra\ (50 hours), as a function of the core radius of the NFW DM profile of the Milky Way, up to the Sun position. Left-hand plot: DM mass of 2.9 TeV (Wino thermal mass). Right-hand plot: DM mass of 9.4 TeV (5plet thermal mass). 
\medskip
}
\end{figure}
\paragraph{Dwarf Spheroidal galaxies versus Galactic Center.} The poor knowledge of the DM density profile, as substantiated in sec.~\ref{sec:GC}, makes it necessary to include this uncertainty when comparing the GC with other targets, like dSphs.
We do so by being agnostic on the size of a possible core radius of the DM distribution in the MW: in fig.~\ref{fig:GC_vs_dSph} we show the projected \CTA\ sensitivities to $\gamma$-lines, from \Tri, \Dra\, and the GC\footnote{
We do not show \Scu\ because of its poorer reach, compared to the other two dSphs. Notice then that in fig.~\ref{fig:GC_vs_dSph} we are comparing observations made by the northern \CTA\ site (the dSphs) with others made by the southern site (the GC).
}
with the assumption of a cored NFW density profile, as a function of the core radius $r_c$. We do so for two representative values of the DM mass, corresponding to the thermal masses of Wino DM and of the MDM 5plet.

One sees in fig.~\ref{fig:GC_vs_dSph} that the sensitivity from 100 hours of GC observations is expected to be comparable with that of 50 hours of observation of a dSph like \Tri\, if the DM distribution develops a core around one kpc, a possibility well allowed by both observations and numerical DM simulations.
For a dwarf like \Dra\ to constitute a better target than the GC, the DM distribution should instead have a core which extends to a few kpc, the mean expected sensitivities of the GC and \Dra\ crossing around 4 kpc. We remind the reader that while observationally such large cores are perfectly viable (and possibly  favored), they are in some tension with most numerical simulations, see sec.~\ref{sec:GC}.

\begin{figure}[!t]
\centering
\includegraphics[width=0.495\textwidth]{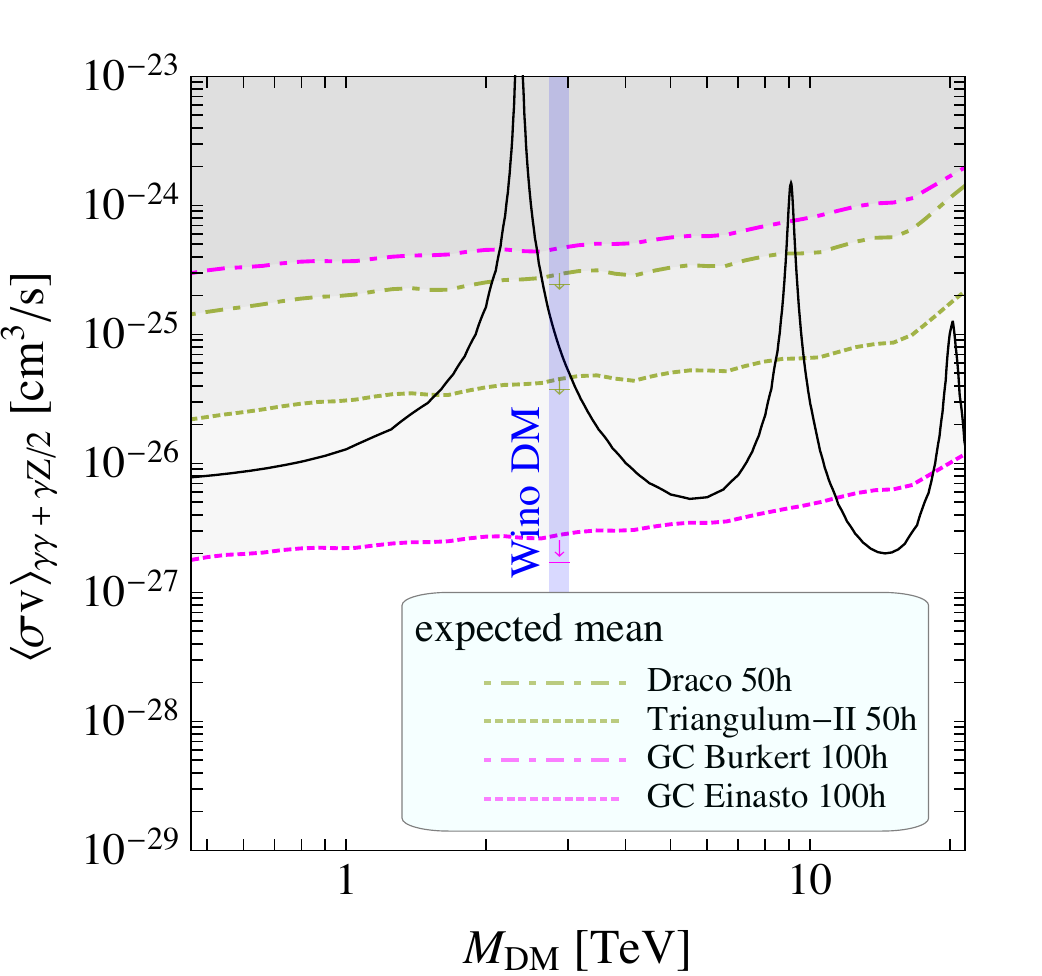}
\includegraphics[width=0.495\textwidth]{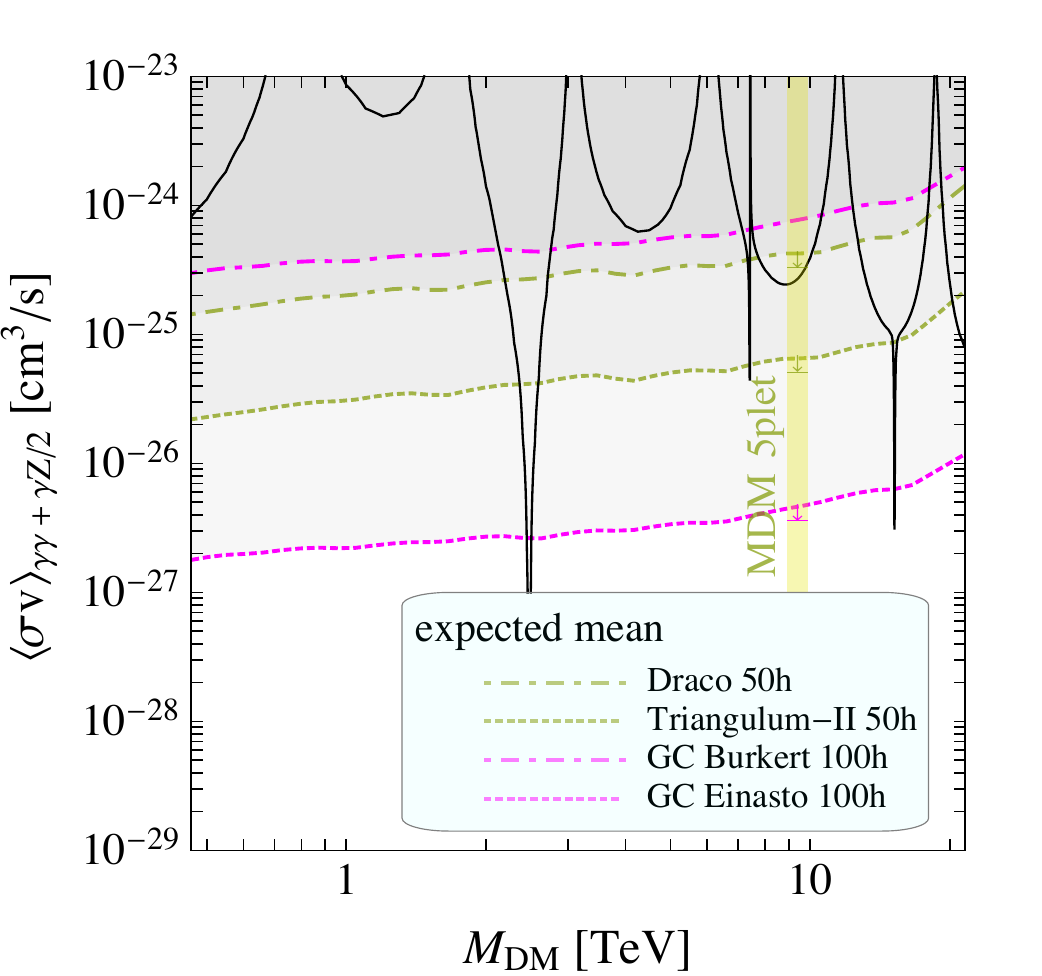} \\
\caption{\em \small \label{fig:CTAProspects}
{\bfseries EW multiplets with CTA.}
Continuous black lines: theoretical prediction of the cross section into monochromatic photons $\langle \sigma v \rangle_{\gamma\gamma + \gamma Z/2}$, for Wino DM (left) and MDM 5plet (right). Overlaid lines: mean expected \CTA\ sensitivities for 50 hours of observation of \Dra\ (dot-dashed ocra) and \Tri (dotted ocra), and for 100 hours of observation of the GC, for a Burkert (dot-dashed magenta) and an Einasto (dotted magenta) profiles.
Vertical shadings as in fig.~\ref{fig:NR_Sommerfeld_XS}. The horizontal lines within the vertical shading represent the improvement in sensitivity of each target, at that mass value, from taking into account the lower energy photon continuum spectra, on top of the $\gamma$-ray line.
\medskip
}
\end{figure}
\paragraph{CTA sensitivity to electroweak multiplets.}
Finally, we show in fig.~\ref{fig:CTAProspects} our mean expected \CTA\ reaches compared with the predictions of the two benchmarks discussed in sec.~\ref{sec:multiplets}, the Wino DM and the MDM 5-plet.
100 hours of observations of the GC would probe the thermal masses of both candidates, if one will be able to firmly exclude the presence of a DM core of several kpc. Concerning dSphs, 50 hours of observation of \Tri\ have the potential to largely exclude, or discover, both thermal candidates. This last statement is however subject to a collection of more kinematical data regarding \Tri, necessary to confirm or disprove its potential for DM indirect detection.
\Dra\ has instead only the potential to marginally test the MDM 5-plet.
The prospects of \CTA\ searches for monochromatic $\gamma$-ray lines, for values of $M_{\rm DM}$ others than the thermal ones, are alse readable off fig.~\ref{fig:CTAProspects}.
Concerning \CTA\ prospects for $\gamma$ lines from the GC, in recent literature they have been given for both Wino~\cite{Ovanesyan:2014fwa,Chun:2015mka} and fiveplet~\cite{Cirelli:2015bda,Garcia-Cely:2015dda} DM. The mild differences with respect to our work are ascribable to the use of previous determinations of \CTA\ sensitivities by those works~\cite{Bergstrom:2012vd,Ibarra:2015tya}, as well as to the choice of different DM profiles.

For the specific thermal mass values, and for the specific predictions of the Wino and fiveplet, we show also the results of a continuum plus line analysis, see secs.~\ref{sec:lines_continuum} and \ref{sec:likelihood}. One sees in fig.~\ref{fig:CTAProspects} that such a model-dependent analysis has the potential to improve the sensitivities by a few tens of percent, with respect to the sensitivities to $\gamma$-ray lines only. We conservatively choose not to include the prospects for this specific analysis in the case of a Burkert profile, because searches for a $\gamma$-ray continuum from the GC have so far required a morphological analysis. This is based on the ON-OFF technique for signal vs background discrimination, which  is only reliable for cuspy  DM profiles~\cite{Abramowski:2011hc,Lefranc:2015pza}.

\section{Summary and conclusions}
\label{sec:conclusion}

In this paper we have provided \CTA\ sensitivities to $\gamma$-ray line features, from the observation of both the GC and, for the first time, of dSphs. We have chosen \Tri, \Dra, and \Scu\ as most promising (according to~\cite{Hayashi:2016kcy}) candidates. Our results can be summarised as follows

\begin{itemize}
\item[$\diamond$] Concerning the analysis, we have performed it with the most updated instrument response functions and cosmic-ray background estimates, provided by \CTA\ in~\cite{CTA:newperformances}. On top of this irreducible background, for the GC we have added the photon diffuse component, obtained by extrapolating the \FERMI\ measurements~\cite{Ackermann:2012pya,Ackermann:2014usa} at higher energies, that we have found to be non-negligible (see fig.~\ref{fig:bkg}). The latest \CTA\ improved estimate concerning background rejections plays a key role in this conclusion. However, the impact of the diffuse photon component on the expected sensitivities is mild, see fig.~\ref{fig:model_indep_pessimistic}.

\item[$\diamond$] Concerning the expected sensitivities, we have compared them with the current available limits from the GC, finding a factor of improvement of a few-to-ten in the TeV range, see fig.~\ref{fig:GC_pastfuture}. Searches of $\gamma$ lines from dSph are today an almost unexplored avenue, and we have shown their potential by comparing their reach with that of the GC, as a function of a putative coring radius of the DM distribution, see fig.~\ref{fig:GC_vs_dSph}.
We have found that the GC becomes a less promising target than \Tri\ if a core extends up to roughly one kpc, and than \Dra\ if a core extends up to a few kpc.
We remind the reader that cores of several kpc appear to be in conflict with most~\cite{DiCintio:2013qxa,Marinacci:2013mha,Tollet:2015gqa} (but not all~\cite{Mollitor:2014ara}) numerical simulations, and are instead perfectly allowed~\cite{Iocco:2015xga} by observations.
We also remark that more tracers are needed to establish \Tri\ as ``the'' target for future observations, in particular because of the small number of tracers (13) on which its study~\cite{Hayashi:2016kcy} (see also\cite{Genina:2016kzg}) is based. We have chosen it nonetheless as an example of the potential of dSphs, because more observations will hopefully increase the number of observed tracers, and because future surveys (such as  Pan-STARRS~\cite{PanSTARRS,Kaiser:2002zz}, DES~\cite{DES,Flaugher:2004vg} and LSST~\cite{LSST,Tyson:2002nh,Hargis:2014kaa}) are expected to discover a plethora of new dwarf satellites of our galaxy.

\item[$\diamond$] Concerning the specific candidates of SUSY Wino and MDM 5plet, we have compared their predictions with our sensitivities in fig.~\ref{fig:CTAProspects}. \CTA\ has the potential to discover or rule out their thermal relic origin, unless the prospects for \Tri\ will be substantially scaled down, and the existence of a several kpc core in the DM distribution will not be excluded.
We have also performed, for the first time, a dedicated analysis that includes the model-dependent information about the lower-energy continuum photon spectrum of such models, for specific thermal mass values. We find that a similar analysis has the potential to increase the cross section reach by a few tens of percent.
\end{itemize}

Our results challenge the widespread claim that the Galactic Center is the best target for the quest for $\gamma$ lines from DM annihilations, and underline the importance of achieving a better knowledge of the DM density profile in the inner kiloparsecs of the Milky Way.
They further motivate dedicating observation time of \CTA\ to dwarf Spheroidal galaxies. 
The analysis presented in this paper also constitutes a valuable complement to physics studies of future high energy colliders, which invoke strong motivation via the search for TeV-mass DM candidates~\cite{Golling:2016gvc}.

\small{
\paragraph{Acknowledgements}
We thank Arianna Di Cintio, Jose O$\tilde{\text{n}}$orbe, Miguel Pato and Aion Viana for useful discussions. P.P. is grateful to the Laboratoire de Physique Th\`eorique et Hautes Energies ({\sc LPTHE}) for hospitality. F.S. is grateful to the Institut d'Astrophysique de Paris ({\sc Iap}) for hospitality, and to the Mainz Institute for Theoretical Physics ({\sc MITP}) 
for hospitality and support.

\medskip

\footnotesize
\noindent Funding and research infrastructure acknowledgements: 
\begin{itemize}
\item[$\ast$] European Research Council ({\sc Erc}) under the EU Seventh Framework Programme (FP7/2007-2013)/{\sc Erc} Starting Grant (agreement n.\ 278234 --- `{\sc NewDark}' project),

\item[$\ast$] {\sc Erc} Advanced Grant 267117 (`{\sc Dark}') hosted by Universit\'e Pierre \& Marie Curie - Paris~6.

\end{itemize}
}

\bibliographystyle{My}
\small
\bibliography{CTA_lines}

%
\end{document}